\documentclass[12pt]{article}
\pdfoutput=1
\usepackage{amsmath,amsfonts,amssymb}
\usepackage{hyperref}
\usepackage{graphicx}
\usepackage{slashed}
\usepackage{epsfig}
\usepackage{psfrag}
\makeatletter\@addtoreset{equation}{section}\makeatother

\usepackage{bbm}

\def\ep{{\epsilon}}
\def\d{{\delta}}

\def\a{{\alpha}}

\def\l{{\lambda}}

\def\G{{\Gamma}}

\def\sg{{\sigma}}
\def\SG{{\Sigma}}

\def\b{{\beta}}

\def\pp{\partial}

\def\be{\begin{equation}}
\def\ee{\end{equation}}
\def\ba{\begin{eqnarray}}
\def\ea{\end{eqnarray}}
\def\bal#1\eal{\begin{align}#1\end{align}}

\def\pic(#1,#2){\begin{figure}[h!]
\begin{center}
  \includegraphics[width=5cm]{#1.png}
  \caption{#2}
 \end{center}
 \end{figure}\\}
 
 \def\mpic(#1,#2){\begin{figure}[h!]
\begin{center}
  \includegraphics[width=7.5cm]{#1.png}
  \caption{#2}
 \end{center}
\end{figure}\\}
 
 \def\bpic(#1,#2){\begin{figure}[h!]
\begin{center}
  \includegraphics[width=10cm]{#1.png}
  \caption{#2}
 \end{center}
\end{figure}\\}
\def\bbpic(#1,#2){\begin{figure}[h!]
\begin{center}
  \includegraphics[width=13cm]{#1.png}
  \caption{#2}
 \end{center}
\end{figure}\\}

\def\spic(#1,#2){\begin{figure}[h!]
\begin{center}
  \includegraphics[width=3cm]{#1.png}
  \caption{#2}
 \end{center}
 \end{figure}\\}

\def\r{\rightarrow}

\def\f {\frac}
\def\ti{\tilde}

\def\ddd{\cdot\cdot\cdot}
\def\no{\nonumber \\}

\def\ep{\epsilon}
\def\r{\rightarrow}

\def\q{\quad}
\def\qq{\quad\quad}
\def\qqq{\quad\quad\quad}

\def\K{{{\rm K\ddot{a}hler}}}

\begin{document}
\bibliographystyle{utphys}
\begin{titlepage}
\thispagestyle{empty}

\begin{flushright}
UT-Komaba-18-5
\end{flushright}
\begin{center}
\noindent{{\textbf{\Large Janus interface in 
two-dimensional 
\\ \vspace{3mm}
 supersymmetric gauge theories}}}\\
\vspace{2cm}
Kanato Goto and Takuya Okuda
\vspace{1cm}

{\it
 University of Tokyo,
 Komaba,
 Tokyo 153-8902, Japan}\\

\vskip 2em
\vskip 2em

\end{center}

\begin{abstract}

We study the Janus interface, a domain wall characterized by spatially varying couplings, in two-dimensional $\mathcal{N}=(2,2)$ supersymmetric gauge theories on the two-sphere.
When the variations of the couplings are small enough, SUSY localization in the Janus background gives an analytic continuation of the sphere partition function.
This directly demonstrates that the interface entropy is proportional to the quantity known as Calabi's diastasis, as originally shown by Bachas et~al.
When the variations are not small, we propose that an analytic continuation of the sphere partition function coincides with the Janus partition function.
We give a prescription for performing such analytic continuation and computing monodromies.
We also point out that the Janus partition function for the equivariant A-twist is precisely the generating function of A-model correlation functions.

\end{abstract}

\end{titlepage}

\newpage

\setcounter{page}{1}
\pagenumbering{arabic}

\tableofcontents
\section{Introduction and summary}

Quantum field theories come with a rich spectrum of operators.
The operators that are most commonly studied are local operators inserted at points in a spacetime manifold.
It is also quite interesting to study non-local operators supported on a subspace of the spacetime.
In this paper we study codimension-one operators in two dimensions.
We will refer to the codimension-one operators as interfaces.
In the literature ``domain walls'' and ``defects'' are other terms commonly used for the same objects.

In particular we will be concerned with the Janus interface.
By definition, the Janus interface is an interface characterized by different values of coupling constants on the two sides of the interface.
See, for example, \cite{Bachas:2001vj,Bak:2003jk,Clark:2004sb,Brunner:2007ur,Brunner:2008fa}.
A way to construct the 2d supersymmetric Janus interface in a general 2d $\mathcal{N}=(2,2)$ was given in the appendix of~\cite{Gaiotto:2009fs}.

An $\mathcal{N}=(2,2)$ superconformal field theory (SCFT) has exactly marginal couplings associated with chiral or twisted chiral deformations.
We will focus on the twisted chiral ones.
Let us recall the relation 
\ba \label{ZS2-Kahler}
Z_{S^2}(t,\overline{t}) = \left(\frac{\ell}{\ell_0}\right)^{c/3} e^{-K(t,\overline{t})} 
\ea
between the two-sphere partition function $Z_{S^2}(t,\overline{t})$ and the K\"ahler potential $K(t,\overline{t})$ on the (twisted chiral) moduli space parametrized by $(t,\overline{t})$.
Here $c$ is the central charge of the 2d $\mathcal{N}=(2,2)$ SCFT, $\ell$~is the radius of the round sphere, and~$\ell_0$ is a reference length scale.
This relation was conjectured in \cite{Jockers:2012dk} and was established in~\cite{Gomis:2012wy,Gerchkovitz:2014gta}.

Our main object of study is the supersymmetric partition function on the two-sphere~\cite{Benini:2012ui,Doroud:2012xw} in the presence of a BPS Janus interface, in conformal and non-conformal cases.
We will construct a supersymmetric field configuration that defines a Janus interface along the equator of~$S^2$, so that the twisted chiral coupling~$t$ takes value $t_{\rm N}$ at the north pole and value~$t_{\rm S}$ at the south pole.
It is natural to expect that the Janus partition function~$Z_\text{Janus}$ is given by the analytic continuation of the sphere partition function:
\begin{equation} \label{Janus-sphere-continuation}
Z_\text{Janus}(t_{\rm N},\overline{t}_{\rm S})  = \left(\text{analytic continuation of }Z_{S^2}(t,\overline{t}) \text{ to }  t=t_{\rm N}\,,  \overline{t}=\overline{t}_{\rm S} \right)\,.
\end{equation}
This expectation can be validated from a relation found in~\cite{Bachas:2016bzn} between the hemisphere partition function \cite{Sugishita:2013jca,Honda:2013uca,Hori:2013ika} and the boundary contribution to the super Weyl anomaly.
Here we demonstrate by a direct localization calculation that this expectation indeed holds.

The reference~\cite{Bachas:2013nxa} showed that the interface entropy $g$ of the Janus interface in an SCFT is given by a particular combination of analytically continued K\"ahler potentials as\begin{equation}\label{g-diastasis}
2\log g = K(t_{\rm N},\overline t_{\rm N})+ K(t_{\rm S},\overline t_{\rm S}) - K(t_{\rm N},\overline t_{\rm S}) - K(t_{\rm S}, \overline t_{\rm N}) \,.
\end{equation}
The combination on the right-hand side is known as Calabi's diastasis~\cite{MR0057000}.
It is invariant under the K\"ahler transformation
\begin{equation} \label{kahler-trans}
K(t,\overline t) \rightarrow K(t,\overline t)  + f(t) + \overline{f}(\overline t) \,,
\end{equation}
where $f$ is an arbitrary holomorphic function and $\overline{f}$ is the complex conjugate.
It was also explained in~\cite{Bachas:2013nxa}, using the boundary states and the spectral flow, that $g$ can be written as%
\footnote{%
Here we identify the supersymmetric partition functions with the corresponding tt$^*$ partition functions.  
Their equality for conformal theories was conjectured in~\cite{Cecotti:2013mba} and proved in~\cite{Bachas:2016bzn}.
See Section~\ref{sec:SUSY-ttstar} for more discussion on this point.
}
\begin{equation}\label{entropy-Janus-sphere}
g^2 = \frac{| Z_\text{Janus}(t_{\rm N},\overline{t}_{\rm S})  |^2  }{Z_{S^2}(t_{\rm N},\overline{t}_{\rm N})Z_{S^2}(t_{\rm S},\overline{t}_{\rm S}) }\, .
\end{equation}
Thus our result~(\ref{Janus-sphere-continuation}) amounts to a demonstration of the entropy-diastasis relation~(\ref{g-diastasis}) by direct SUSY localization.
This was the original motivation for this work.

We find that the result of localization calculation for $Z_\text{Janus}(t_{\rm N},\overline{t}_{\rm S}) $ depends only on the values of $t$ at the north and the south poles.
We expect that this is an artifact of our localization procedure, and that with a more refined treatment the partition function should depend on the homotopy class of the path in the moduli space along which the moduli $t$ vary as we move from the north pole to the south pole.
Here we take a pragmatic approach and propose that in a correct version of the relation~(\ref{Janus-sphere-continuation}) captures the dependence on the homotopy class.

The K\"ahler potential can be expressed in terms of the periods of the mirror Calabi-Yau manifold. 
There is huge literature on the analytic continuation of such periods.
In this paper we take an alternative approach and offer a prescription based on the expression of the sphere partition function as a residue integral found in~\cite{Jockers:2012dk}.
The integrand involves holomorphic and anti-holomorphic copies of a certain function, which is proportional to Givental's $I$-function, itself a generalization of the hypergeometric function.
(See for example~\cite{MR1354600,1996alg.geom..3021G,MR1653024}.)
The analytic continuation of the holomorphic function can be done by the Mellin-Barnes method.

We organize the paper as follows.
In Section~\ref{sec:janus-conformal} we review general aspects of a Janus interface in $\mathcal{N}=(2,2)$ supersymmetric theories.
In Section~\ref{sec:Janus-GLSM} we specialize to gauged linear sigma models (GLSM's) and perform localization calculations.
We will explain how to perform analytic continuation in Section~\ref{sec:analytic-continuation}.
In Sections~\ref{sec:CY-hypersurface} and~\ref{sec:SQED}, we will study concrete examples and perform explicit calculations.
In Section~\ref{sec:A-twist} we construct a Janus interface on $S^2$ in a theory with equivariant A-twist~\cite{Benini:2012ui,Closset:2015rna}  and observe that the partition function with the interface is a generating function of the A-model correlation functions.
We will conclude with discussion in Section~\ref{sec:discussion}.
Appendices contain useful formulas and details of computations.

\section{Janus interface in 2d  $\mathcal{N}=(2,2)$ supersymmetric theories}\label{sec:janus-conformal}

\subsection{Symmetries}

We begin by studying the symmetries of $\mathcal{N}=(2,2)$ supersymmetric theories and their interfaces.
We will consider non-conformal as well as conformal cases.

A general $\mathcal{N}=(2,2)$ theory in Minkowski space has four Poincar\'e supercharges $Q_\pm$ and $\overline{Q}_\pm$ as we review in Appendix~\ref{sec:SUSY-Minkowski}.
They satisfy the relations $(Q_\pm)^\dagger =\overline{Q}_\pm$.
An interface extending along the time direction preserves the Hamiltonian while it breaks the translation generated by the momentum.
The interface we will study preserves the combination $Q_+ + e^{-i\beta}Q_-$ and its hermitian conjugate $\overline{Q}_+ + e^{i\beta}\overline{Q}_-$.%
\footnote{%
Here $\beta$ is real, but in Euclidean signature it can be complex.  In that case the combinations cease to be the hermitian conjugate of each other.
}
The symmetry group in Lorentzian signature is generated by the hermitian combinations of generators.

Some of the supersymmetric models we consider flow to an $\mathcal{N}=(2,2)$ SCFT.
The supersymmetry algebra of the bulk theory is enhanced to two copies of $\mathcal{N}=2$ super Virasoro algebra.
Let us perform the Euclidean rotation to the complex plane, and also geometrically rotate by angle $\pi/2$ so that the interface extends in the spatial direction.
The superconformal algebra has left-moving generators usually denoted as $L_m, J_m, G^\pm_r$ and their right-moving counterparts $\widetilde L_m, \widetilde J_m, \widetilde G^\pm_r$.
See~\cite{Lerche:1989uy,Recknagel:1997sb} for notations.
The global part, which preserves the Riemann sphere, of the superconformal group is generated by~$L_m$ ($m=0,\pm 1$), $J_0$, $G^\pm_r$ ($r=\pm 1/2$), and their right-moving partners.
The complexification of the $\mathcal{N}=(2,2)$ superconformal algebra is therefore $sl(2|1,\mathbb{C}) \oplus sl(2|1,\mathbb{C})$.%
\footnote{%
Note also that $sl(2|1,\mathbb{C}) \simeq osp(2|2,\mathbb{C})$. 
}
The usual conformal transformation maps the Euclidean plane to an infinite cylinder $S^1\times\mathbb{R}$.
Rotating back to Lorentzian signature with $S^1$ taken as space, hermitian conjugation acts as $L_m^\dagger = L_{-m}$, $(G^\pm_r )^\dagger = G^\mp_{-r}$, etc.
Again the symmetry group in Lorentzian signature is generated by the hermitian combinations.
An interface wrapping $S^1$ can be regarded as an operator $\mathcal{I}$ on the Hilbert space.
A B-type superconformal interface satisfies
\begin{equation} \label{interface-super-Virasoro}
[ L_m - \widetilde{L}_{-m} ,\mathcal{I}] 
=
[ J_m +\widetilde{J}_{-m} ,\mathcal{I}] 
=
[ G_r^\pm  + i \eta  \widetilde{G}_{-r}^\pm ,\mathcal{I}] 
=
0\, ,
\end{equation}
where the sign $\eta =\pm 1$ specifies the spin structure.%
An A-type superconformal interface satisfies the same relations with the replacement $\widetilde{J}_m \rightarrow -\widetilde{J}_m$, $\widetilde{G}^\pm_r \rightarrow \widetilde{G}^\mp_r$ (mirror automorphism).
We expect that in conformal theories, the supersymmetric interfaces we study flow to superconformal interfaces.

We explained the characterization~(\ref{interface-super-Virasoro}) of the interfaces to make contact with the literature.
In this paper we will mainly study theories on $S^2$.

On $S^2$, one can couple a possibly non-conformal $\mathcal{N}=(2,2)$ theory with vector (or axial) R-symmetry $u(1)_\text{V}$ (or $u(1)_\text{A}$) to a version of supergravity that contains a gauge field for that R-symmetry.%
\footnote{%
We write $u(1)_\text{V}$ (Lie algebra) rather than $u(1)_\text{V}$ (Lie group) because the R-charges are not quantized on the supersymmetric sphere, where the total R-symmetry flux vanishes.
}
A suitable supergravity background with a round metric preserves four fermionic symmetries that generate $su(2|1)_\text{V}$  (or $su(2|1)_\text{A}$).
The interface we will consider breaks the symmetry group to $u(1|1)_\text{V}$ (or $u(1|1)_\text{A}$).

A different supergravity background~\cite{Gomis:2012wy} with a deformed metric (\ref{deformed-sphere-metric}) only preserves $u(1|1)_\text{V}$ (or $u(1|1)_\text{A}$).
In this case an interface does not break symmetries further.

\subsection{Off-shell construction of the Janus interface}\label{sec:off-shell-construction}

In two dimensions, a conformal field theory with $\mathcal{N}=(2,2)$ supersymmetry comes with a moduli space $\mathcal{M}$ of exactly marginal couplings.
The  space $\mathcal{M}$ is locally a product:  $\mathcal{M}\simeq \mathcal{M}_\text{c}\times \mathcal{M}_\text{tc}$.
The factors $ \mathcal{M}_\text{c}$ and $ \mathcal{M}_\text{tc}$ are parametrized by the coupling constants that can be viewed as scalar components of chiral and twisted chiral supermultiplets, respectively, in accordance with the philosophy of~\cite{Seiberg:1993vc}.
In this paper we focus on the physics associated with the twisted chiral part $ \mathcal{M}_\text{tc}$.

The Janus interface preserving B-type supersymmetry was defined in~\cite{Gaiotto:2009fs}, extending the similar constructions in 4d~\cite{Bak:2003jk,Clark:2004sb,Gaiotto:2008sd}.
It is characterized by a profile of the coupling $t\in\mathcal{M}_\text{tc}$ that depends on the position in the spacetime, in such a way that B-type supersymmetry is preserved.

A basic idea of this paper is that we can construct a supersymmetric Janus interface in the off-shell supersymmetry formalism by promoting coupling constants to supermultiplets and turning on auxiliary fields.
The values of the auxiliary fields are fixed by the condition that the supersymmetry variations of the fermionic partners vanish.
Concretely, let us consider a twisted chiral multiplet  $T=(t,\chi^T_-,\overline{\chi}_-^T,E^T)$ in Minkowski space, where $t(x^1)$ is the coupling, $\chi_-$ and $\overline{\chi}_-$ are fermions, and $E$ is the auxiliary field.
The metric is
\ba
ds^2=  - dx^+ dx^-\, ,
\ea
where $x^\pm = x^0 \pm x^1$.
We are interested in the B-type supersymmetry specified by the condition on the SUSY parameters $(\epsilon_\pm,\overline{\epsilon}_\pm)$
\ba \label{B-type-parameters}
\ep_+=-e^{-i\b}\ep_- \,,
\quad
\overline{\ep}_+=- e^{+i\b}\overline{\ep}_- \,,
\ea
where $\beta$ is a real constant.
The vanishing of the fermions and their variations
\begin{equation}
\d\chi_-^T=i\ep_+ (\partial_1-\partial_0)t+\ep_-E^T \,, 
\quad
\d\overline{\chi}_+^T=i\overline{\ep}_- (\partial_1+\partial_0)t+\overline{\ep}_+E^T \, 
\end{equation}
implies that
\begin{equation} \label{E-v-relation}
E^T = i e^{-i\beta} \partial_1 t \,.
\end{equation}
In Minkowski signature complex conjugation gives $\overline{E}{}^T = -i e^{i\beta} \partial_1 \overline{t}$ while in Euclidean signature we have the same relation but $\beta$ can be complex.

We will explain the off-shell construction even more explicitly for GLSM's in Section~\ref{sec:Janus-Minkowski-GLSM}.

\subsection{SUSY sphere partition function and tt$^*$ partition function}\label{sec:SUSY-ttstar}

\begin{figure}[tbp]
\centering
\includegraphics[width=13cm]{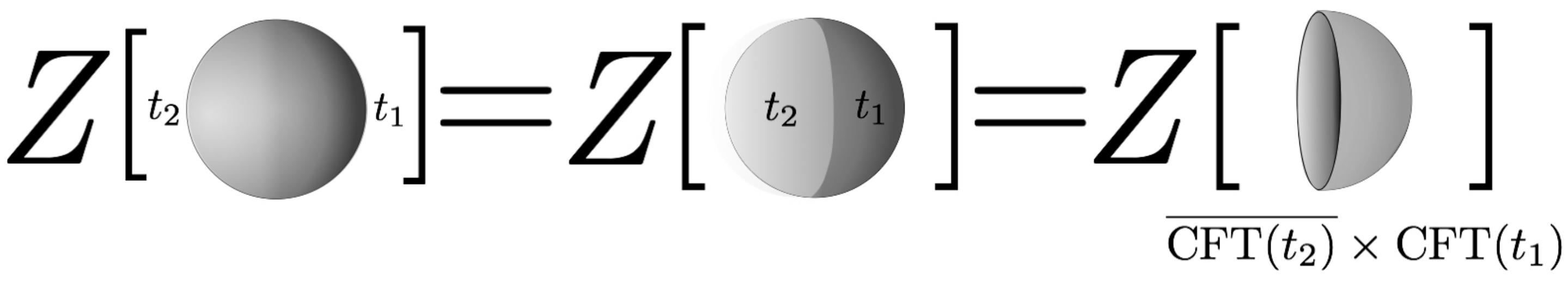}
\caption{\label{figure:folding}%
The expectation value of the Janus interface with a smooth profile of couplings coincides with that of zero width.
Upon folding, it is then identified with the hemisphere partition with a particular boundary condition for a product theory.
}
\end{figure}

In general a Janus interface can have a finite width, meaning that the couplings vary over a region of finite width.
The supersymmetric sphere partition function $Z_\text{Janus} $ in the presence of the Janus interface is, however, independent of the width of the interface.
This is because a change in the profiles of the couplings fixing the values at the two poles results in a SUSY-exact shift of the action~\cite{Gaiotto:2009fs}.%
\footnote{%
See footnote~\ref{footnote:action-exact}.
}
Then we can make it infinitely thin without changing the value of  $Z_\text{Janus} $.
We expect that in this limit we obtain a superconformal interface.

Let us consider folding the configuration, turning the thin interface into a boundary with a B-type boundary condition for a product theory.
See Figure~\ref{figure:folding}.
It was conjectured in~\cite{Cecotti:2013mba} that the supersymmetric hemisphere partition function for a B-type boundary condition coincides with the $tt^*$ hemisphere partition function.
The conjecture was proved in~\cite{Bachas:2016bzn} by establishing a relation between the hemisphere partition function and the boundary contribution to the super Weyl anomaly.%
\footnote{%
The authors of~\cite{Bachas:2016bzn} claim that they also compute Calabi's diastasis using localization.
Indeed they show in their paper that a hemisphere partition function depends holomorphically on the bulk moduli.
Upon conformal perturbation in the bulk, the hemisphere partition function of the folded theory can then only depend holomorphically on $t_{\rm N}$ and anti-holomorphically on $t_{\rm S}$, leading to the diastasis for an appropriate ratio of partition functions.
Thus their approach is a combination of SUSY localization for the pure sphere partition function and conformal perturbation theory for the Janus interface.
In this paper we use explicit localization for the latter part as well.
We thank C.~Bachas for a discussion and clarification on this point.
}
Since the equality of the supersymmetric and $tt^*$ partition functions holds in the presence of an interface, the same also holds for the partition functions with the trivial interface, or in other words for the supersymmetric and tt$^*$ sphere partition functions.

\section{Janus interface in gauged linear sigma models}\label{sec:Janus-GLSM}

We now study the Janus interface reviewed above in the specific setting of $\mathcal{N}=(2,2)$ gauged linear sigma models, which we review in Appendix~\ref{sec:GLSM-review}.
In particular we present a construction that makes use of the off-shell formulation of $\mathcal{N}=(2,2)$ supersymmetry.

To avoid cluttering equations, we start with a single $U(1)$ factor.
We will generalize the results of the analysis to multiple $U(1)$ factors when writing down the formula for the Janus partition function.

\subsection{Janus configuration in Minkowski space}\label{sec:Janus-Minkowski-GLSM}

The terms in the action that involve the Fayet-Iliopoulos (FI) parameter $r$ and the theta angle $\vartheta$ are%
\begin{equation}
S_\text{FI-$\vartheta$}= - \frac{1}{4\pi}\int d^2xd^2\widetilde \theta\, t\SG + c.c. =\frac{1}{2\pi}\int d^2x (-rD +\vartheta v_{01}).
\end{equation}
Here we applied (\ref{twisted-chiral-superpotential-Minkowski}) to $\widetilde{W} = - (t/4\pi)\Sigma$.%

Let us promote the complexified FI parameter $t$ to a twisted chiral superfield%
\begin{equation}
\begin{aligned}
T=t(\tilde{y}) + \theta^+ \overline{\chi}{}^T_+(\tilde{y})  +\overline{\theta}{}^- \chi{}^T_-(\tilde{y}) +\theta^+\overline{\theta}{}^- E{}^T(\tilde{y})  \,,
\end{aligned}
\end{equation}
where $\tilde{y}^\pm = x^\pm \mp i\theta^\pm \overline{\theta}{}^\pm$.
We consider the twisted superpotential
\begin{equation} \label{twisted-superpotential-FI-theta}
\widetilde{W}_\text{Janus} = - \frac{1}{4\pi} T\Sigma 
\end{equation}
with the superfield coupling $T$.
To preserve supersymmetry we impose the conditions $\delta \chi^T=\delta \overline{\chi}^T=0$ and $\chi^T=\overline{\chi}^T=0$ which give the relation~(\ref{E-v-relation}).
We obtain
\begin{equation}
\begin{aligned}
S_\text{Janus}
&:=
-\frac{1}{4\pi}\int d^2xd^2\widetilde\theta\, T \SG+ c.c. \label{S-Janus-twisted-superpotential}
\\
&=\frac{1}{2\pi}\int d^2x \left[- r(x^1) D + \vartheta(x^1) v_{01} \right]+\frac{i}{4\pi}\int  d^2x \left[e^{-i\b}(\pp_1t)\sigma-e^{i\b} (\pp_1\overline{t})\overline{\sigma}\right]\,.
\end{aligned}
\end{equation}
The SUSY variation $\delta$ can be expressed in terms of supercharges as 
$$
i\epsilon_+ {Q}_- - i\epsilon_- Q_+  - i\overline\epsilon_+ \overline{Q}_- + i \overline \epsilon_- \overline{Q}_+ \,.
$$
The condition~(\ref{B-type-parameters}) corresponds to the B-type combination $Q_+ + e^{-i\beta}Q_-$ and its hermitian conjugate.
The action can then be written as
\begin{equation}
S_\text{Janus} = \frac{1}{4\pi} \int dx^0 dx^1
\left(
t(x^1) \{ Q_+, [\overline{Q}_-,\sigma]\} + i e^{-i\beta} (\partial_1 t)\sigma + \text{h.c.}
\right)\, .
\end{equation}
In this form the equivalence with the on-shell construction of the Janus interface as given in~(A.4) of~\cite{Gaiotto:2009fs} is manifest.%

As noted in~\cite{Gaiotto:2009fs}, and as follows from the first line of~(\ref{S-Janus-twisted-superpotential}), the variation of the action under $t(x^1) \rightarrow t(x^1) + \Delta t(x^1)$ is $\delta$-exact if $ \Delta t (x^1)\rightarrow 0$ as~$|x^1|\rightarrow  \infty$.%
\footnote{%
Thus when the system is placed on $S^1\times\mathbb{R}$, slightly different profiles with the same asymptotic values of $t$ act  in the same way on Ramond-Ramond ground states on $S^1$.
As explained in~\cite{Gaiotto:2009fs}, this implies the flatness of the $tt^*$ connection, and hence all the $tt^*$ equations~\cite{Cecotti:1991me}.
}
This implies that when we study quantities protected by SUSY we can continuously deform the profile $t(x^1)$ while fixing the asymptotics, {\it as long as the theory remains regular during the deformation process.}
If the theory becomes singular, for example due to the non-compactness in the field space at the conifold singularity discussed later, the physical quantities can jump discretely.
Thus the BPS physics depends on the homotopy class of the path $t(x^1)$ in the moduli space.

We may choose $\Delta t$ so that $t(x^1)$ becomes a step function~\cite{Gaiotto:2009fs}
\ba
t(x^1)=\begin{cases}
t_-& x^1<0\\
t_+& x^1>0
\end{cases}
\ea
with two constants $t_\pm$.
The delta function in $\pp_1t=(t_+-t_-)\d(x^1)$ gives
\ba
S_{{\rm Janus}}=\frac{1}{2\pi}\int d^2x ( - r D + \vartheta v_{01})+\frac{i}{4\pi}\int_{x^1=0}  dx^2 (e^{-i\b}(t_+-t_-)\sigma-e^{i\b}(\overline{t}_+-\overline{t}_-)\overline{\sigma}) \,.
\ea
We write the first term in a way identical to the usual FI and theta couplings, but the values of $(r,\vartheta)$ on the two sides of the thin interface are different.
In addition, we get a contribution involving the scalar $\sigma$ localized along the interface.

\subsection{Janus configuration on the supersymmetric sphere}

Let us apply the off-shell construction to the supersymmetric sphere~\cite{Benini:2012ui,Doroud:2012xw}.
We work with the deformed sphere metric \cite{Gomis:2012wy}
\begin{equation} \label{deformed-sphere-metric}
ds^2= {\rm f}(\theta)^2 d\theta^2 + \ell^2 \sin^2\theta d\varphi^2 
\end{equation}
and the vielbein $e^{\hat{z}}= e^{i\varphi} (d\theta + i \ell \sin\theta d\varphi)$, $e^{\hat{\overline z}}= e^{-i\varphi} (d\theta - i \ell \sin\theta d\varphi)$.
The function ${\rm f}(\theta)>0$ behaves as ${\rm f} =\ell +\mathcal{O}(\theta^2)$ near $\theta=0$ and as ${\rm f} =\ell +\mathcal{O}((\pi-\theta)^2)$ near $\theta=\pi$ respectively. In the case of the round sphere, we have ${\rm f}(\theta)=\ell$. The SUSY parameters are
\begin{equation}
(\ep_-,\ep_+)=(e^{-i\varphi}\sin\frac{\theta}{2}, - \cos\frac{\theta}{2}) \,,
\quad
(\overline{\ep}_-,\overline{\ep}_+)=(-\cos\frac{\theta}{2}, e^{ i\varphi}\sin\frac{\theta}{2}) \,.
\end{equation}

We are interested in a configuration of the complexified FI parameter 
\begin{equation}
t(\theta) = r(\theta) -i \vartheta(\theta)
\end{equation}
 that depends only on the longitudinal coordinate $\theta$.
To preserve SUSY,  the fermionic partners of $t, \overline{t}$ and their variations are set to zero.
Formulas in (\ref{SUSY-twisted-chiral}) and (\ref{SUSY-twisted-anti-chiral}) give
\begin{equation}
E^T =\frac{i}{{\rm f}(\theta)}\cot \f\theta2 
\cdot
 \pp_\theta t(\theta) \,, \quad
\overline{E}{}^T = -\frac{i}{{\rm f}(\theta)}\tan\f\theta2 \cdot \pp_\theta \overline{t}(\theta)  \,.
\end{equation}
Note that $E^T$ and $\overline{E}{}^T$ are not the complex conjugate of each other.

Localization in the Janus background proceeds as in \cite{Benini:2012ui, Doroud:2012xw,Gomis:2012wy}.
The saddle point configurations are given by
\ba
\sigma= \sigma^{(0)}  - i\frac{B}{2\ell } \,, \qquad v_\varphi =
 \frac{B}{2}(1-\cos\theta) \,.
\ea
We need to integrate over the constant $\sigma^{(0)}$ and sum over the magnetic flux $B\in\mathbb{Z}$.

\subsection{Sphere partition function with a Janus interface}

We now turn to a Janus interface on a two-sphere, where the moduli vary along the longitudinal direction.

\subsubsection{Classical action}\label{sec:classical-action}

We can evaluate the on-shell classical action $S_\text{cl}$ by the formula (\ref{twisted-F-term-coupling-rigid}) with the twisted superpotential (\ref{twisted-superpotential-FI-theta}).

To evaluate the classical action in the presence of a Janus interface, it is useful to notice that the twisted superpotential action (\ref{twisted-F-term-coupling-rigid}) can be written as%
\footnote{\label{footnote:action-exact}%
Note that (\ref{twisted-F-term-coupling-rigid}) with $\widetilde{W}=v$, $\overline{\widetilde{W}}=\overline{v}$ reduces to $S=- \int \left[E -  (i/f) v + \overline{E} -   (i/f) \overline{v} \right]$.
The formulas for the fermion valuations allow us to write $S=-4\pi i \ell (v|_{\theta=0} + \overline{v}|_{\theta=\pi}) + \delta(\ldots)$.

The result~(\ref{on-shell-tilde-W-exact}) implies that the Janus partition function for a general $\mathcal{N}=(2,2)$ theory (not necessarily a gauged linear sigma model) depends only on the twisted chiral parameter $t_\text{N}$ at the north pole and the  twisted anti-chiral parameter $\overline{t}_\text{S}$ at the south pole.
}
\begin{equation} \label{on-shell-tilde-W-exact}
S_{\widetilde{W}} = - 4\pi i \ell \big(  \widetilde{W} |_{\theta=0} + \overline{\widetilde{W}} |_{\theta=\pi}  \big) +\delta(\ldots)\,. 
\end{equation}
in the supersymmetric background.
A similar expression appeared earlier, for example in \cite{Closset:2015rna} for the sphere with equivariant A-twist.
The result is that the on-shell action for the Janus interface is obtained by replacing $(r,\theta)$ with the effective FI and theta parameters
\begin{equation}\label{eff}
\begin{aligned}
\underline{r} 
&:=\frac{r_{\rm N}+r_{\rm S}}{2}-i\frac{\vartheta_{\rm N}-\vartheta_{\rm S}}{2}  \,, \\
\underline{\vartheta}
 &:=\frac{\vartheta_{\rm N}+\vartheta_{\rm S}}{2}+i\frac{r_{\rm N}-r_{\rm S}}{2} \,,
\end{aligned}
\end{equation}
where the subscripts ${\rm N}$ and ${\rm S}$ indicate that the symbols are evaluated at the north and  south poles respectively.
Thus we can write the classical action in the presence of a Janus interface as
\begin{equation}
\begin{aligned}
S_\text{cl} (\text{Janus}) & =  + 2 i \underline{r}  \ell  \sigma^{(0)}  + i B \underline{\vartheta} 
\\
&= t_{\rm N} \left(  + i \ell \sigma^{(0)}  -  \frac{B}{2} \right) + \overline{t}_{\rm S} \left(  +  i \ell \sigma^{(0)}  +  \frac{B}{2} \right)
\,.
\end{aligned}
\end{equation}

\subsubsection{Janus partition function} \label{sec:Janus-general}

Suppose that the gauge group is $\prod _{a=1}^{s}U(1)_a$ and that the chiral multiplet $\Phi_j$ has charge~$Q^a_j$ under $U(1)_a$.
We use the notation ${\boldsymbol B} =(B_a)$ for magnetic fluxes.
Let $q_j$ be the vector R-charge of $\Phi_j$.
Recall from the above that we need to integrate over the constant mode $\sigma^{(0)}$.
In the following we use the notation~${\boldsymbol \sigma}=( \sigma_a)$,  where $\sigma_a$ denotes the combination $ - i \ell \sigma^{(0)}$ for each $U(1)_a$.
The usual localization procedure as in~\cite{Benini:2012ui, Doroud:2012xw} gives the Janus partition function written as
\bal
Z_\text{Janus}&=\sum_{\boldsymbol B} 
\int_C
\prod_{a=1}^{s} \frac{d^s {\boldsymbol \sigma}}{(2\pi i)^s} 
Z_{{\rm cl}} ({\boldsymbol \sg}, {\boldsymbol B}  )
 \prod_j Z_j ({\boldsymbol \sigma},{\boldsymbol B} )\, ,
\eal
where 
\begin{equation}
Z_{{\rm cl}} ({\boldsymbol \sg}, {\boldsymbol B}  )=
 \exp\left[t_{\rm N}^a \left(\sigma_a  + \frac{B_a}{2}\right)
+ \overline{t}{}_{\rm S}^a \left(\sigma_a  - \frac{B_a}{2}\right)
 \right]
 \,,\\
\end{equation}
and
\begin{equation}
Z_j ({\boldsymbol \sigma},{\boldsymbol B} )=  \frac{\displaystyle \Gamma\Big(\frac{q_j}{2}  + Q_j^a \, \sg_a  -   \frac{1}{2} Q^a_j \, B_a \Big)}
{\displaystyle \Gamma\Big(1-\frac{q_j}{2} -  Q_j^a \, \sg_a  - \frac{1}{2} Q_j^a B_a \Big)}\,.
\end{equation}
The symbols $t_{\rm N}^a$ and $\overline{t}{}_{\rm S}^a$ denote the values of the renormalized and complexified FI parameters.
The contour $C$ of integration needs to be taken appropriately for given such parameters.

This reduces to the usual sphere partition function~\cite{Benini:2012ui, Doroud:2012xw} if we set $t_{\rm N} \rightarrow t$, $\overline{t}{}_{\rm S} \rightarrow \overline{t}$.
In other words, the Janus partition function is obtained simply by analytic continuation $t \rightarrow t_{\rm N}$, $\overline{t} \rightarrow\overline{t}{}_{\rm S} $.
Thus the sphere partition function for a GLSM  with a Janus interface 
is given by the analytic
continuation of the sphere partition function via
\begin{equation} \label{Z-Janus-Z-sphere}
 Z_\text{Janus} =  Z_{S^2}(t_{\rm N},\overline{t}{}_{\rm S}) \,.
\end{equation}
When the GLSM reduces a conformal field theory, this gives the analytic continuation of the K\"ahler potential:
\begin{equation} \label{Z-Janus-K}
 Z_\text{Janus} =\left(\frac{\ell}{\ell_0}\right)^{c/3} e^{-K(t_{\rm N},\overline{t}{}_{\rm S})} \,.
\end{equation}
As in~(\ref{ZS2-Kahler}) we included the prefactor that depends on the central charge $c$ given by $c/3=\sum_j(1-q_j)-s$ of the IR CFT;
it arises from a more careful regularization and renormalization, independently of the presence of the interface.
 
\subsubsection{K\"ahler transformations for the analytically continued K\"ahler potential}
For the usual sphere partition function of an $\mathcal{N}=(2,2)$ conformal theory, a finite supergravity counterterm accounts for the K\"ahler transformation (\ref{kahler-trans}) with $t$ and $\overline{t}$ being the complex conjugate of each other \cite{Gerchkovitz:2014gta}.
A component expression for the counterterm was given in~\cite{Closset:2014pda}.
Let~$R$ be the Ricci scalar, $F^\text{R}_{\mu\nu}$ the vector R-symmetry gauge field, and $\mathcal{H}$ the graviphoton field strength.
The counterterm in the current convention (see~\cite{Okuda:2017rwo}) is
\begin{equation} \label{counterterm-Kahler-trans}
\begin{aligned}
S_{f,\overline{f}}
&=\frac{1}{4\pi} \int d^2x \sqrt{g} \Big[\epsilon^{\mu\nu} F^\text{R}_{\mu\nu} \left(f(t)-\overline{f}(\overline{t})\right) +  \frac{R}{2}\left( f(t) +\overline{f}(\overline{t})\right)
\\
&\qquad\qquad\qquad\qquad\qquad\qquad\qquad
-\mathcal{H} E^T\partial_t f(t)
-\overline{\mathcal{H}} \,\overline{E}{}^T\partial_{\overline{t}} \overline{f}(\overline{t}) 
 \Big] \,,
 \end{aligned}
 \end{equation}
where we have set the fermions to zero.
We introduced an arbitrary holomorphic function~$f(t)$ and its complex conjugate $\overline{f}(\overline{t})$.
The same trick as in Section \ref{sec:classical-action} works here;~(\ref{counterterm-Kahler-trans}) can be written as
$ f(t)|_{\theta=0} + \overline{f}(\overline{t})|_{\theta=\pi} +\delta(\ldots) $.
The counterterm then evaluates to
\begin{equation} \label{finite-counterterm-evaluated}
S_{f,\overline{f}}
= f(t_{\rm N}) + \overline{f}( \overline{t}_{\rm S}) \,.
\end{equation}
The addition of the counterterm has the effect
\begin{equation}
Z_\text{Janus}(t_{\rm N},\overline{t}{}_{\rm S}) \rightarrow e^{ - f(t_{\rm N}) - \overline{f}( \overline{t}_{\rm S}) }Z_\text{Janus}(t_{\rm N},\overline{t}{}_{\rm S}) \,,
\end{equation}
which is equivalent to the K\"ahler transformation (\ref{kahler-trans}) with the replacement $t\rightarrow t_{\rm N}$, $\overline{t}\rightarrow \overline{t}_{\rm S}$.

\section{Analytic continuation and monodromy}\label{sec:analytic-continuation}

The relation between the space of complexified FI parameters $(t^a)$ and the moduli space~$ \mathcal{M}_{\rm tc}$ of exactly marginal twisted chiral couplings involves identifications by $t^a \sim t^a + 2\pi i$ and monodromies around singularities.
As mentioned in the introduction, the localization result~(\ref{Z-Janus-Z-sphere}) as well as the value of the counterterm~(\ref{finite-counterterm-evaluated}) depend only on the values of twisted chiral parameters at the north and south poles,  not on the homotopy class of the path $\gamma$ given by the map $ [0,\theta] \rightarrow \mathcal{M}_{\rm tc}$, $\theta \mapsto [t(\theta)]$, where $[\,\bullet\,]$ denotes the equivalence class defined by the identifications above.

In the rest of the paper we take the pragmatic approach and assume that the Janus partition function is given by the analytic continuation along $\gamma$.
There are codimension-one singular loci in $ \mathcal{M}_\text{tc}$, and it is an interesting problem to study the effect of the monodromy along a closed curve $\gamma$ that begins and ends at the large volume limit after going around singular loci.

Our computation will involve a certain function $\Psi_\lambda(x)$ that naturally arises in the analysis of Picard-Fuchs equations for GLSM's by the Frobenius method.
This function coincides with the so-called $I$-function~\cite{MR1354600,1996alg.geom..3021G,MR1653024} up to an $x$-independent but $\lambda$-dependent multiplicative factor.
We will make contact with~\cite{Knapp:2016rec,Erkinger:2017aaa} where analytic continuation and monodromy in GLSM was studied in terms of the hemisphere partition function.

For a general $\mathcal{N}=(2,2)$ SCFT,  the moduli space $\mathcal{M}_\text{tc}$ of exactly marginal twisted chiral couplings is a local (or projective) special K\"ahler manifold \cite{MR717607,Strominger:1990pd,Freed:1997dp}.
In particular $\mathcal{M}_\text{tc}$ admits local projective coordinates $X^I$ and a homogeneous function $\mathcal{F}(X^I)$ of degree two, {\it i.e.}, $\mathcal{F}(\lambda X^I)=\lambda^2 \mathcal{F}(X^I)$.
In terms of these data, the K\"ahler potential is given by
\begin{equation} \label{K-symplectic-basis}
e^{-K} = i (\overline{X}{}^I \mathcal{F}_I - X^I \overline{\mathcal{F}}_I) \,,
\end{equation}
where 
\begin{equation}
 \mathcal{F}_I  =\frac{\partial \mathcal{F}}{\partial X^I} \,.
\end{equation}
When we view $X^I(t)$ and $\mathcal{F}_I(t)$ as holomorphic sections on $\mathcal{M}_\text{tc}$ dependent on local coordinates $t=(t^a)$, the K\"ahler transformation (\ref{kahler-trans}) corresponds to the holomorphic rescaling $X^I(t) \rightarrow e^{-f(t)} X^I(t)$.
We may write the expectation value of the Janus interface as
\begin{equation} \label{Janus-symplectic-combination}
Z_\text{Janus}  =  i  \mathcal{F}_I(t_{\rm N})\overline{X}{}^I(\overline t_{\rm S}) - i X^I(t_{\rm N}) \overline{\mathcal{F}}_I(\overline{t}_{\rm S})  \,.
\end{equation}

\section{Calabi-Yau hypersurface in a projective space}\label{sec:CY-hypersurface}

In this section we study a specific class of GLSM's.
The gauge group is $U(1)$, and the theory has chiral multiplets $\Phi_i$ ($i=1,\ldots,n$) and $P$.
We assign gauge and vector R-charges as follows.
\begin{alignat}{3}
&\qqq &U(1) \q & u(1)_\text{V}
\no
&\Phi_i &+1 \qq &2q
\no
&P &-n \qq &2-2nq
\nonumber
\end{alignat}
We take the superpotential
\begin{equation}
W = P \cdot G(\Phi)\,,
\end{equation}
where $G(\Phi)$ is a homogeneous polynomial of degree $n$.
Let us denote the bottom components of $\Phi_i$ by $\phi_i$.
We assume that the equations 
\ba
G=\frac{\pp G}{\pp \phi_1}=\ddd=\frac{\pp G}{\pp \phi_n}=0
\ea
have no solution except the trivial one $\phi_1=\ddd=\phi_n=0$. The equation
\ba
G(\phi_1,\ddd, \phi_n)=0
\ea 
defines a complex hypersurface $M$ in $\mathbb{CP}^{n-1}$.  
For $r\gg 0$, it is known that this theory flows to the  $\mathcal{N}=(2,2)$ non-linear sigma model with the Calabi-Yau $M$ as its target space. 
The complexified Kahler class of $M$ is identified as $t$ in this limit. 
For $r\ll 0$ this theory reduces to the Landau-Ginzburg orbifold.

\subsection{Janus partition function}\label{sec:hypersurface-Janus-PF}
We set $\ell=\ell_0$.
The sphere partition function without an interface is~\cite{Benini:2012ui, Doroud:2012xw}\begin{equation}
\label{eq: hypersurface-sigma-int}
Z_{S^2}=\sum_{B\in\mathbb{Z}}e^{-i B \vartheta}
\int^{i\infty}_{-i\infty}
\frac{d\sigma}{2\pi i}e^{ 2 r\sigma}\biggl(\frac{\Gamma(q  + \sigma-\frac{B}{2})}{\Gamma(1-q  - \sigma-\frac{B}{2})}\biggl)^n\frac{\Gamma(1- n q   -  n \sigma  + n \frac{B}{2})}{\Gamma(n q  + n  \sigma + n \frac{B}{2})} \,.
\end{equation}
The integration contour asymptotes to $\pm i \infty$, and is chosen to separate the poles $\sigma_{\rm p}$ of the first factor in the integrand (${\rm Re}(\sigma_{\rm p} +q)\leq 0$)  from those of the second factor  (${\rm Re}(\sigma_{\rm p} +q)\geq 1/n$).
We consider the region $r\gg0$ where the theory flows in the IR to the non-linear sigma model with a target space $M$. 
In this region, the contour can be closed in the left half-plane yielding the answer%
\ba \label{eq: hypersurface-mod-squared}
Z_{S^2}(t,\overline{t})=(x \overline{x})^q\oint\frac{d\ep}{2\pi i}(x \overline{x})^{-\ep}\frac{\pi^{n-1}\sin(n\pi \ep)}{\sin^n(\pi \ep)}\biggl|
\sum_{k=0}^{\infty} x^k\frac{\Gamma(1+nk-n\ep)}{\Gamma(1+k-\ep)^n}\biggl
|^2 
\ea
with the notation $| F(x,\epsilon)|^2 := F(x,\epsilon) \overline{F}(\overline{x},\epsilon)$ and $x:=e^{-t+n\pi i}=e^{-r+i(\vartheta+n\pi)}$, $\overline{x}:=e^{-\overline{t}-n\pi i}=e^{-r-i(\vartheta+n\pi i)}$.
The $\epsilon$-integration contour is a small circle around the origin.

Next let us consider the sphere partition function with an interface
\begin{equation} \label{hypersurface-Janus-integral}
Z_\text{Janus}=\sum_{m\in\mathbb{Z}}e^{-im\underline{\vartheta}}\int_{C}
\frac{d\sigma}{2\pi i}e^{ 2 \underline{r}\sigma}\biggl(\frac{\Gamma(q  + \sigma-\frac{B}{2})}{\Gamma(1-q  - \sigma-\frac{B}{2})}\biggl)^n\frac{\Gamma(1- n q   -  n \sigma  + n \frac{B}{2})}{\Gamma(n q  + n  \sigma + n \frac{B}{2})} \,.
\end{equation}
One issue is the choice of integration contour $C$.
For $\pm {\rm Im}\,\sigma \gg 0$, the integrand behaves as $\exp[(r_{\rm N}+r_{\rm S}-i \vartheta_{\rm N}+i \vartheta_{\rm S} \mp  2n\log n)\sigma]=: \exp[ c\, e^{i\alpha_\pm } \sigma]$ with $c\gg 0$ and $\alpha_\pm \in \mathbb{R}$.
When $\vartheta_{\rm N} =  \vartheta_{\rm S}$ the phase $\alpha_\pm $ vanishes and the integral strictly along the imaginary axis converges.%
\footnote{%
A convenient summary of the conditions for such convergence is Lemma~3.3 of \cite{Horja-hypergeometric}.
}
When $\vartheta_{\rm N}\neq  \vartheta_{\rm S}$, we have non-zero phases $\alpha_\pm \neq 0$ and the integral strictly along the imaginary axis is no longer oscillatory; for convergence we need to tilt the contour $C$ so that it is contained in the region of the $\sigma$-plane where the integral converges.
With an appropriate choice of the contour, we conclude that
\begin{equation}
\begin{aligned}
Z_\text{Janus}(t_{\rm N},\overline{t}_{\rm S}) &= (x_{\rm N} \overline{x}_{\rm S})^q\oint\frac{d\ep}{2\pi i}(x_{\rm N} \overline{x}_{\rm S})^{-\ep}\frac{\pi^{n-1}\sin(n\pi \ep)}{\sin^n(\pi \ep)}
\\
&
\times 
\biggl(\sum_{k=0}^{\infty}  x_{\rm N}^k\frac{\Gamma(1+nk-n\ep)}{\Gamma(1+k-\ep)^n}\biggl)\biggl(\sum_{k=0}^{\infty}  \overline x_{\rm S}^k\frac{\Gamma(1+nk-n\ep)}{\Gamma(1+k-\ep)^n}\biggl) \,.
\end{aligned}
\end{equation}
This gives the analytic continuation $K(t_{\rm N},\overline{t}_{\rm S})$ of the $\K$ potential via
\ba
Z_{\rm Janus}(t_{\rm N},\overline{t}_{\rm S})=e^{-K(t_{\rm N},\overline{t}_{\rm S})}.
\ea

\subsection{Function $\Psi_\lambda(x)$} \label{sec:Psi-lambda}

Let us define
\begin{equation} \label{Psi-lambda-def}
\Psi_\lambda(x) :=
\sum_{k=0}^\infty \frac{\Gamma(1+ n(k+\lambda))}{\Gamma(1+k+\lambda)^n} x^{k+\lambda}  
\end{equation}
and
\begin{equation} \label{Phi-j-def}
\Phi_j(x):= \frac{1}{(2\pi i)^j j!} \left.\frac{\partial^j}{\partial \lambda^j} \Psi_\lambda(x) \right|_{\lambda=0} \,,
\qquad j=0,\ldots, n-2 \,.
\end{equation}
By Frobenius' argument, these give a basis of solutions to the Picard-Fuchs equation
 \begin{equation} \label{Picard-Fuchs}
\left[
(x\partial_x)^{n-1} - n^n x \prod_{j=1}^{n-1} (x \partial_x + j/n) \right] \Phi(x)  =0 \,.
\end{equation}

Function $\Psi_\lambda(x)$ is proportional to the so-called $I$-function~\cite{MR1354600,1996alg.geom..3021G,MR1653024} up to an $x$-independent but $\lambda$-dependent multiplicative factor.%
\footnote{%
The precise relation between $\Psi_\lambda(x)$ and the $I$-function $I$ \cite{1996alg.geom..3021G} is
\begin{equation}
I(-t,\lambda) = \frac{\Gamma(1+\lambda)^n}{\Gamma(1+n\lambda)} \Psi_\lambda(e^{-t}) \,.
\end{equation}
The $I$-function was studied using the supersymmetric sphere partition function in~\cite{Bonelli:2013mma}.
}
The flat coordinate usually denoted as $B+iJ=\tau$ is given by $\partial_\lambda \Psi_\lambda/(2\pi i\Psi_\lambda)$ evaluated at $\lambda=0$:
\begin{equation} \label{flat-coord-hypersurface}
\begin{aligned}
\tau
&= \frac{1}{2\pi i}
\frac{
\displaystyle
 \sum_{k=0}^\infty \frac{\Gamma(1+ n k)}{\Gamma(1+k)^n}
\left(
\log x
+ n \psi(1+nk) -n \psi(1+k)
\right)
x^k
}{
\displaystyle
\sum_{k=0}^\infty \frac{\Gamma(1+ n k)}{\Gamma(1+k)^n}
x^k
}
\,.
\end{aligned}
\end{equation}
Here $\psi(z)=\Gamma'(z)/\Gamma(z)$ denotes the digamma function.

\subsection{Monodromies}

\label{sec:monodromies}

We can write the sphere partition function (\ref{eq: hypersurface-mod-squared}) as
\ba \label{eq: hypersurface-mod-squared2}
Z_{S^2}(t,\overline{t})=(x \overline{x})^q\oint\frac{d\ep}{2\pi i}\frac{\pi^{n-1}\sin(n\pi \ep)}{\sin^n(\pi \ep)}
\Psi_{-\epsilon}(  
x
) \Psi_{-\epsilon}( 
 \overline{x}
 ) \,.
\ea
Thus we can obtain the monodromy transformation of $Z_{S^2}(t,\overline{t})$ from the knowledge of the monodromy of $\Psi_\lambda(x)$.

In the $x$-plane, the only singularities are at $x=0$ (large volume), $n^{-n}$ (conifold point), and $\infty$ (Landau-Ginzburg orbifold point).
Under the shift $\vartheta \rightarrow \vartheta + 2\pi i$ ($x\rightarrow e^{2\pi i} x$) around $x=0$, there is a monodromy
\begin{equation} \label{}
\Psi_\lambda(x)
\rightarrow
e^{2\pi i \lambda} \Psi_\lambda(x)
\,.
\end{equation}
In Appendix~\ref{app:analytic-continuation} we show that analytic continuation along the loop that goes around the singularity at $x=n^{-n}$ counterclockwise gives
\begin{equation} \label{Psi-monodromy-conifold}
\Psi_\lambda(x)
\rightarrow
\Psi_\lambda(x)
-
 \oint_{0} d\mu
\frac{  1 - e^{-2\pi in \mu}}{(1- e^{-2\pi i \mu})^n}
\Psi_\mu(x) 
\,.
\end{equation}
The $\mu$-integration contour goes around $\mu=0$ counterclockwise.
Loops around $x=0$ and $x= n^{-n}$ generate all possible non-trivial loops.

\subsection{Cubic elliptic curve}

We set $x= e^{-(t-3\pi i)} $.
Let us introduce $\Phi_0$ and $\Phi_1$ by
\begin{equation}
\Psi_\lambda(x) = \Phi_0(x) +2\pi i  \lambda \Phi_1(x) + \mathcal{O}(\lambda^2) \,.
\end{equation}
The flat coordinate is
\begin{equation}
B+i J = \tau = \frac{\Phi_1(x)}{\Phi_0(x)} \,.
\end{equation}

The Janus partition function is given in terms of $\Phi_0$ and $\Phi_1$ by
\begin{equation}
Z_\text{Janus} (t_{\rm N}, \overline t_{\rm S}) 
= -6 \pi i  (x_{\rm N} \overline x_{\rm S})^q (
\Phi_0 (x_{\rm N}) \Phi_1 (\overline x_{\rm S})
+
\Phi_0 (\overline x_{\rm S}) \Phi_1 (x_{\rm N})
)\, .
\end{equation}
We note that $\overline{\Phi_1(x)} = - \Phi_1( \overline{x})$.
Thus we have the relation
\begin{equation}
\frac{|Z_\text{Janus} (t_{\rm N}, \overline t_{\rm S}) |^2}{
Z_{S^2} (t_{\rm N}, \overline t_{\rm N})
Z_{S^2} (t_{\rm S}, \overline t_{\rm S})
 }
 =
 \frac{
( \tau_{\rm N} - \overline{\tau}_{\rm S} ) ( \tau_{\rm S} - \overline{\tau}_{\rm N} )
 }{
 ( \tau_{\rm N} - \overline{\tau}_{\rm N} ) ( \tau_{\rm S} - \overline{\tau}_{\rm S} )
 }
 \,.
\end{equation}
The interface entropy $g$ is given by
\begin{equation}
g=\sqrt{ \frac{
( \tau_{\rm N} - \overline{\tau}_{\rm S} ) ( \tau_{\rm S} - \overline{\tau}_{\rm N} )
 }{
 ( \tau_{\rm N} - \overline{\tau}_{\rm N} ) ( \tau_{\rm S} - \overline{\tau}_{\rm S} )
 }
}
\,,
\end{equation}
which reproduces a basic example of~\cite{Bachas:2013nxa}.

Let us study the monodromies.
Under  the minimal shift of $\vartheta$ for $r\gg 0$, {\it i.e.}, $x \rightarrow e^{2\pi i} x$, we have the monodromy
\begin{equation}
\begin{pmatrix}
\Phi_1
\\
\Phi_0 
\end{pmatrix}
\rightarrow
\begin{pmatrix}
\Phi_1 + \Phi_0
\\
\Phi_0 
\end{pmatrix}
=
\begin{pmatrix}
1 & 1
\\
0 & 1
\end{pmatrix}
\begin{pmatrix}
\Phi_1
\\
\Phi_0 
\end{pmatrix} \,.
\end{equation}
When going around the singularity $x=3^{-3}$ counterclockwise, the monodromy is given by
\begin{equation}
\begin{pmatrix}
\Phi_1
\\
\Phi_0 
\end{pmatrix}
\rightarrow
\begin{pmatrix}
\Phi_1 
\\
\Phi_0  - 3\Phi_1
\end{pmatrix}
=
\begin{pmatrix}
1 & 0
\\
-3 & 1
\end{pmatrix}
\begin{pmatrix}
\Phi_1
\\
\Phi_0 
\end{pmatrix} \,.
\end{equation}
These transformations  act on $\tau$ by M\"obius tranformations.

They do not generate the full modular group $SL(2,\mathbb{Z})$.%
\footnote{%
See, for example, \cite{Hosono:2000eb}.
}
Rather, they generate  what is known as the congruence subgroup $\Gamma_1(3)$ of $SL(2,\mathbb{Z})$.
It is defined as the group of $SL(2,\mathbb{Z})$ matrices~$\begin{pmatrix}
a & b\\
c & d
\end{pmatrix}
$ with $a\equiv d\equiv 1$, $c\equiv 0$ (mod 3).
This means that in our gauged linear sigma model description, the cubic elliptic curve and its mirror carry extra structures in addition to the complexified K\"ahler and complex structures, respectively, which are preserved by the monodromies visible in the description.
Let us recall that the T-duality group of the non-linear sigma model with torus target is $PSL(2,\mathbb{Z}) \times PSL(2,\mathbb{Z}) \rtimes (\mathbb{Z}_2)^2$.
Its subgroup that acts on the complexified K\"ahler modulus $\tau$ holomorphically is $PSL(2,\mathbb{Z})$.
The monodromy group $\Gamma_1(3)$ visible in the current gauged linear sigma model description is only a subgroup of the $SL(2,\mathbb{Z})$, whose quotient by its center is $PSL(2,\mathbb{Z})$.
In particular, the values of $t$ that correspond to $\tau$ and $-1/\tau$ give the identical IR theory%
\footnote{%
For example $x_0=0.00181623$ corresponds to the point $\tau=i$ invariant under ${\bf S}: \tau \rightarrow -1/\tau$.
Two points near $\tau =i$ related by ${\bf S}$ correspond to two distinct points near $x_0$, which are not related by~$\Gamma_1(3)$.
}; this equivalence is not manifest in the gauged linear sigma model description.

\begin{figure}[htbp]
\centering
\includegraphics[scale=0.9]{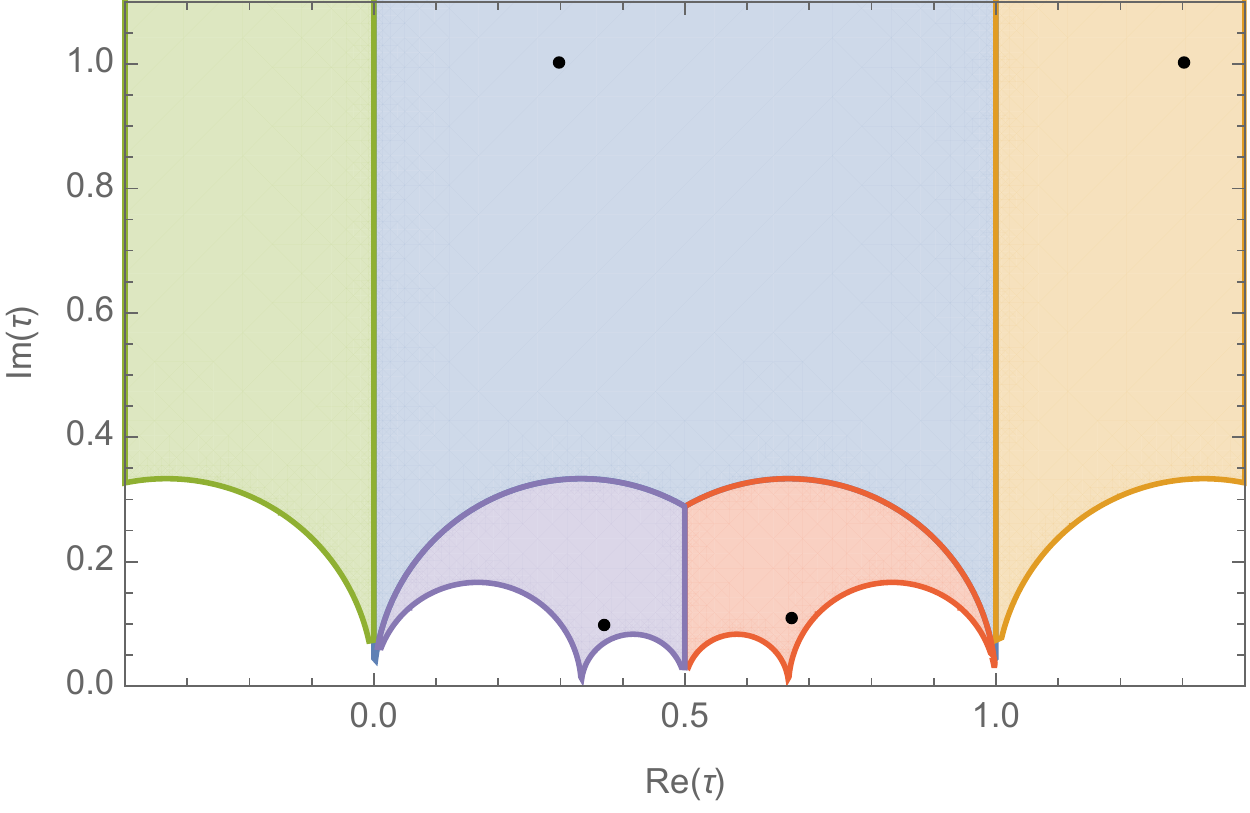}
\caption{\label{figure:fund-domain}%
Copies of the fundamental domain of $\Gamma_1(3)$ by its action.
The dots are the copies of the point $0.3+ i$.
}
\end{figure}

The fundamental domain of $\Gamma_1(3)$ is shown in Figure~\ref{figure:fund-domain}.
The singularity $x=3^{-3}$ is mapped to the cusp at $\tau=0$, where the volume of the elliptic curve is 0, and in the mirror elliptic curve a 1-cycle shrinks.
The Landau-Ginzburg point $x=\infty$ is mapped to the corners $\tau= \pm \frac{1}{2} +i \frac{\sqrt{3}}{6}$.
See~\cite{Jockers:2006sm} for a related analysis.

\subsection{Quintic} \label{sec:quintic}

This is a specialization of the above to $n=5$.
We set $x=e^{-t+5\pi i}$.
Specializing (\ref{flat-coord-hypersurface}) to $n=5$, we have the flat coordinate
\begin{equation} \label{flat-coord-quintic}
B+i J =\tau =
\frac{-(t - 5\pi i) - 770 e^{-t} + \mathcal{O}(e^{-2t})}{2\pi i}
\,.
\end{equation}

\subsubsection{Monodromy in quantum K\"ahler moduli space}

While $\Phi_j$ in~(\ref{Phi-j-def}) provide a basis of solutions to the Picard-Fuchs equation~(\ref{Picard-Fuchs}), another useful basis is provided by hemisphere partition functions.
Let us compute the monodromy matrices in this basis and compare with the literature for consistency check.
The hemisphere partition function $Z_{\mathcal{B}} (t) $ for a D-brane $\mathcal{B}$ can be expressed, in the geometric phase, as
\begin{equation} \label{Z-hem-f}
\begin{aligned} 
Z_{\mathcal{B}} (t) &=\int_{i\mathbb{R}}\frac{d\sigma}{2\pi i}e^{t \sigma}\Gamma(\sigma)^{5}\Gamma(1-5\sigma) f_{\mathcal{B}}(e^{2\pi i \sigma}) 
\\
&=
\oint_0 \frac{d\epsilon}{2\pi i} \frac{f_{\mathcal{B}}(e^{2\pi i\epsilon})}{\sin^5(\pi \epsilon)} e^{5\pi i \epsilon} \Psi_{-\epsilon}(x) 
\,,
\end{aligned}
\end{equation}
where the ``brane factor'' $f_{\mathcal{B}}(y) \in \mathbb{Z}[y,y^{-1}]$, a Laurent polynomial with integer coefficients, is determined by the D-brane $\mathcal{B}$.
Note that $Z_{\mathcal{B}}$ depends only on the equivalence class of $f_{\mathcal{B}}(y)$ in $\mathbb{Z}[y,y^{-1}]/\langle (y-1)^5 \rangle$, where $\langle (y-1)^5 \rangle$ is the ideal generated by $  (y-1)^5$.
As explained in~\cite{Hori:2013ika}, (\ref{Z-hem-f}) is equivalent to the central charge formula conjectured in~\cite{Hosono:2004jp}.

We make  the following choices for $f_{\mathcal{B}}$~\cite{Knapp:2016rec}:
\begin{center}
\begin{tabular}{c|c}
brane & $f_{\mathcal{B}}(y)$
\\
\hline
D0 & $(y-1)^4 $ 
\\
D2 & $y^{-1}(y^{-1}-1)^3 $ \\
D4 & $(1-y)(y^{-5}-1)$ \\
D6 & $y^{-5}-1$
\end{tabular}
\end{center}
We find
\begin{equation} \label{Z-hem-Phi}
\begin{aligned}
 \pi Z_\text{D0} &= 16   \Phi_0(x)\,,\\
\pi Z_\text{D2} &=  16 \Phi_1(x)\,, \\
\pi Z_\text{D4} &=    80    \Phi_0(x) -40 \Phi_1(x)+  80 \Phi_2(x) \,, \\
\pi  Z_\text{D6} &=  \frac{200 }{3  } \Phi_1(x)  + 80 \Phi_3(x) \,.
   \end{aligned}
\end{equation}
The analytic continuation of $\Phi_j(x)$ in $x$ can be obtained by taking derivatives, as in (\ref{Phi-j-def}), of the analytically continued $\Psi_\lambda(x)$.
We can use (\ref{Z-hem-Phi}) to read off the monodromy action on the hemisphere partition functions.

The monodromy for the clockwise loop around the conifold point, as follows from (\ref{Psi-monodromy-conifold}), acts as
\begin{equation} \label{quintic-monodromy-conifold}
\begin{pmatrix}
Z_\text{D0}  \\
Z_\text{D2}  \\
Z_\text{D4}  \\
Z_\text{D6}  \\
\end{pmatrix}
\rightarrow
\begin{pmatrix}
1 & 0 & 0 & -1 \\
 0 & 1 & 0 & 0 \\
 0 & 0 & 1 & -5 \\
 0 & 0 & 0 & 1 \\
\end{pmatrix}
\begin{pmatrix}
Z_\text{D0}  \\
Z_\text{D2}  \\
Z_\text{D4}  \\
Z_\text{D6}  \\
\end{pmatrix}
\,.
\end{equation}

For the loop in the large volume $\vartheta\rightarrow \vartheta + 2\pi$ ($x\rightarrow e^{2\pi i} x$), we similarly find
\begin{equation} \label{quintic-monodromy-large-volume}
\begin{pmatrix}
Z_\text{D0}  \\
Z_\text{D2}  \\
Z_\text{D4}  \\
Z_\text{D6}  \\
\end{pmatrix}
\rightarrow
\begin{pmatrix}
1 & 0 & 0 & 0 \\
 0 & 1 & 0 & 0 \\
 0 & 5 & 1 & 0 \\
 0 & 5 & 1 & 1 \\
\end{pmatrix}
\begin{pmatrix}
Z_\text{D0}  \\
Z_\text{D2}  \\
Z_\text{D4}  \\
Z_\text{D6}  \\
\end{pmatrix} \,.
\end{equation}

The monodromy matrices in (\ref{quintic-monodromy-conifold}) and (\ref{quintic-monodromy-large-volume}) are in agreement with~\cite{Knapp:2016rec}.
They generate the full monodromy group on the quantum K\"ahler moduli space of the quintic~\cite{Candelas:1990rm}.

Since D-branes provide states in the Hilbert space of string theory, the D-brane hemisphere partition functions form a natural integral basis.
This is why the monodromy matrices above are integral.
The sphere partition function (and hence the Janus partition function) can be factorized, {\it i.e.}, written as a sum of products of a hemisphere partition function and the complex conjugate of another.
The coefficients that appear in the factorization are the inverse of the matrix of Dirac indices (or Mukai pairings), which are conjecturally identified with the cylinder partition functions specified by two boundary conditions~\cite{Honda:2013uca,Hori:2013ika}.
The cylinder partition functions provide a natural symplectic pairing.

For the prepotential of the quintic~\cite{Candelas:1990rm}%
\footnote{%
Our overall sign convention for the prepotential is that of~\cite{Candelas:1990rm}, and is the opposite of \cite{Hosono:1994ax,Jockers:2012dk}.
This ensures that the expression (\ref{K-symplectic-basis}) for  $e^{-K}$ is positive in the large volume limit.
The coefficients of the polynomial part other than $\tau^3$ are subject to ambiguities due to monodromy $\tau \rightarrow \tau+1$.
}
\begin{equation}
\mathcal{F}(X^0,X^1=X^0 \tau)= (X^0)^2\left( - \frac{5}{6} \tau^3 - \frac{11}{4} \tau^2 + \frac{25}{12} \tau - \frac{25 i}{2\pi ^3} \zeta(3) + \mathcal{O}(e^{2\pi i \tau}) \right) \,,
\end{equation}
the symplectic basis in Section~\ref{sec:analytic-continuation} is related to the D-brane basis as
\begin{equation}
\begin{pmatrix}
X^0 \\ X^1 \\ \mathcal{F}_1 \\ \mathcal{F}_0
\end{pmatrix}
=
\begin{pmatrix}
1 & 0 & 0 & 0 \\
 0 & 1 & 0 & 0 \\
 5 & -8 & -1 & 0 \\
 0 & 0 & 0 & 1 \\
\end{pmatrix}
\begin{pmatrix}
Z_\text{D0}  \\
Z_\text{D2}  \\
Z_\text{D4}  \\
Z_\text{D6}  \\
\end{pmatrix} \,.
\end{equation}
The Janus partition function is then given by (\ref{Janus-symplectic-combination}).

\section{SQED}\label{sec:SQED}

In this section we study the Janus interface in a non-conformal model, the SQED.
We also compute the monodromies of the partition functions and compare them with those obtained in~\cite{Gomis:2014eya}.

It was found in~\cite{Doroud:2012xw,Gomis:2014eya} that the sphere partition function of the $\mathcal{N}=(2,2)$ $U(N)$ gauge theory with $N_{\rm F}$ fundamental and $N_{\rm F}$ anti-fundamental chiral multiplets coincides with an $A_{N_{\rm F}-1}$ Toda CFT four-point function up to a constant factor that can be identified with the (deformed) four-sphere partition function~\cite{Hama:2012bg} of $N_{\rm F}^2$ free hypermultiplets.
SQED corresponds to the special case $N=1$.

SQED has $N_{\rm F}$ chiral multiplets of gauge charge $+1$ and the same number of chiral multiplets of charge $-1$.
The theory has flavor symmetry $\left( U(N_{\rm F})\times U(N_{\rm F}) \right) /U(1)$, where the quotient is by the gauge group.
We can couple the theory to background gauge multiplets for the flavor symmetry and give vevs (twisted masses) to the scalar components.
We assume that the positively charged fields have twisted masses ${\rm m}_f$ and R-charges $q_f$, while the negatively charged ones have twisted masses $\tilde {\rm m}_f$ and R-charges~$\tilde{q}_f$.
For non-zero twisted masses, the theory ceases to be conformal.
In this section we use the notations $m_f:= i \ell {\rm m}_f + q_f/2$ (not to be confused with magnetic fluxes in Section~\ref{sec:Janus-general}), $\tilde{m}_f:= - i\ell \tilde{\rm m}_f - \tilde{q}_f/2$.%
\footnote{%
The definitions in this paper and in~\cite{Gomis:2014eya} differ by factors $\pm i$: $m_f^{\rm here}=-im_f^{\rm there}$, $\tilde{m}_f^{\rm here}=i\tilde{m}_f^{\rm there}$.}
The final result will be invariant under the simultaneous shifts $m_f \rightarrow m_f + a_0$, $\tilde{m}_f \rightarrow \tilde{m}_f + a_0$ by the common constant $a_0$, reflecting the quotient in flavor group.

\subsection{Janus partition function }

In the  Coulomb branch representation, the sphere partition function of SQED  is given as%
\footnote{%
Relative to 
the expression in~\cite{Gomis:2014eya}, (\ref{S2-PF-SQED}) has the opposite sign in $e^{-i B \vartheta}$.
The formula~(\ref{S2-PF-SQED}) is actually invariant under $ \vartheta \rightarrow - \vartheta$ because of the identity $\Gamma(x-B/2)/\Gamma(1-x-B/2)=(-1)^B\Gamma(x+B/2)/\Gamma(1-x+B/2)$ for $B\in\mathbb{Z}$, the presence of an even number of chiral multiplets with charges $\pm 1$, and the sum over $B\in\mathbb{Z}$.
}
\begin{equation}\label{S2-PF-SQED}
Z^{\rm SQED}_{S^2}=\sum_{B\in\mathbb{Z}}
e^{-i B \vartheta}
\int^{+i\infty}_{-i\infty}
\f{d\sigma}{2\pi i}
e^{2r\sigma}
\prod_{f =1}^{N_{\rm F}}\f{\Gamma(m_f+\sigma-\f{B}{2})}{\Gamma(1-m_f-\sigma-\f{B}{2})}\f{\Gamma(-\tilde{m}_f-\sigma+\f{B}{2})}{\Gamma(1+\tilde{m}_f+\sigma+\f{B}{2})}\, .
\end{equation}
The contour of integration asymptotes to $\pm i\infty$ and is taken to pass to the right of $-m_f$ and to the left of $-\tilde{m}_f$ for all $f\in\{1,\ldots,N_\text{F}\}$.
We impose the condition $\sum_f {\rm Re}(m_f-\tilde{m}_f)<N_\text{F}$ on the R-charges to ensure that the integral converges.

Let us define variables $x:= e^{-t + N_{\rm F} \pi i} = e^{-r + i (\vartheta + N_{\rm F} \pi)}$, $\overline{x}:= e^{- \overline{t} - N_{\rm F} \pi i} = e^{-r - i (\vartheta + N_{\rm F} \pi)}$.
For $|x|<1$ we can close the contour of the integration in the left half plane.
Manipulations similar to those in~Section \ref{sec:hypersurface-Janus-PF} lead to
\bal
Z^{\rm SQED}_{S^2}
&=\oint_C\f{d\l }{2\pi i}
\Bigg(
\prod_{f=1}^{N_{\rm F}}\f{\sin\pi(\l-\ti{m}_f)}{\sin\pi (\l-m_f)}
\Bigg)
\Psi_{\l}^{({\rm s})}(
x
)\Psi_{\l}^{({\rm s})}(
\overline{x}
),\label{Zs}\eal
where  we defined%
\footnote{%
The superscripts $({\rm s})$ and $({\rm u})$ refer to the crossing channels for the Toda four-point function~\cite{Gomis:2014eya}.
}
\begin{equation}\label{Zsp}
\Psi_{\l}^{({\rm s})}(x):=\sum_{k=0}^\infty x^{k+\l}
\prod_{f=1}^{N_{\rm F}} \f{\Gamma(k+\l-\tilde{m}_f)}{\Gamma(k+1+\l-m_f)}  \,.
\end{equation}
The contour $C$ encloses all the poles at $\l=m_f$, $f\in \{1,\ldots,N_{\rm F}\}$.
Our proposal outlined in Section~\ref{sec:analytic-continuation} is that the partition function in the presence of a Janus interface is given by the analytic continuation
\begin{equation}\label{SQED-Janus-s-channel}
Z_\text{Janus}^\text{SQED}(t_\text{N}, \overline{t}_\text{S})
=\oint_C\f{d\l }{2\pi i}
\Bigg(
\prod_{f=1}^{N_{\rm F}}\f{\sin\pi(\l-\ti{m}_f)}{\sin\pi (\l-m_f)}
\Bigg)
\Psi_{\l}^{({\rm s})}( x_\text{N} )
\Psi_{\l}^{({\rm s})}( \overline{x}_\text{S}) 
\end{equation}
of the sphere partition function.

Similarly, for $|x|>1$ we obtain
\begin{equation}\label{Zu}
Z^{\rm SQED}_{S^2}
=\oint_{C'}\f{d\zeta }{2\pi i}
\Bigg(
\prod_{f=1}^{N_{\rm F}}\f{\sin\pi(\zeta-m_f)}{\sin\pi (\zeta-\ti{m}_f)}
\Bigg)
\Psi_{\zeta}^{({\rm u})}(
x
)\Psi_{\zeta}^{({\rm u})}(
\overline{x}
)
\end{equation}
for the sphere partition function, where
\begin{equation} \label{Zup}
\Psi_{\zeta}^{({\rm u})}(x): =\sum_{k=0}^\infty x^{-k+\zeta}\prod_{f=1}^{N_{\rm F}}\f{\Gamma(k+m_f-\zeta)}{{\Gamma(k+1+\tilde{m}_f-\zeta)}}\, .
\end{equation}
The contour $C'$ encloses all the poles at $\zeta=\ti{m}_f$, $f\in \{1,\ldots,N_{\rm F}\}$.

\subsection{Analytic continuation and monodromies}

For $0<   {\rm arg}\,x < 2\pi$,  the functions $ \Psi^{({\rm s})}_\l(x)$ and $\Psi^{({\rm u})}_{\xi}(x)$ are related as 
\begin{equation}\label{relesu2}
\begin{aligned}
  \Psi^{({\rm s})}_\l(x)&=-\oint_{C_1}\f{d\xi}{2\pi i}
  \Bigg(
  \prod_{f=1}^{N_{\rm F}}\f{\sin\pi(\xi-m_f)}{\sin\pi(\xi-\ti{m}_f)}
  \Bigg)
  \f{(-1)^{N_{\rm F}}\pi e^{-\pi i (\xi-\l)}}{\sin\pi (\xi-\l)}\Psi^{({\rm u})}_{\xi}(x)\, ,
\\
\Psi_{\l}^{({\rm u})}(x)&=-\oint_{C_2}\f{d\xi}{2\pi i}
  \Bigg(
  \prod_{f=1}^{N_{\rm F}}\f{\sin\pi(\xi-\ti{m}_f)}{\sin\pi(\xi-m_f)}
  \Bigg)
  \f{(-1)^{N_{\rm F}}\pi e^{-\pi i (\xi-\l)}}{\sin\pi (\xi-\l)}\Psi^{({\rm s})}_{\xi}(x)\,.
\end{aligned}
\end{equation}
The contour $C_1$  encloses the poles $\{\ti{m}_f | f=1,\ldots,N_\text{F}\} \cup \{\l-1\}$ (and no others), while $C_2$  encloses $\{m_f |  f= 1,\ldots,N_\text{F}\}\cup \{\l+1\}$.
We derive (\ref{relesu2})  in Appendix~\ref{sec:SQED-Psi}.

From the expressions (\ref{Zs}) and (\ref{Zu}), we can obtain the monodromy transformations of the sphere partition function $Z^{\rm SQED}_{S^2}$ through the monodromies of $\Psi_{\l}^{({\rm s})}(x)$ and $\Psi_{\lambda}^{({\rm u})}(x)$.

The $x$-plane has three singularities at $x=0$, $1$, and $\infty$.
Under the counterclockwise rotation around $x=0$ $(x\r e^{2\pi i}x)$, we have the monodromy 
\ba
\Psi_{\l}^{({\rm s})}(x)\r e^{2\pi i \l}\Psi_{\l}^{({\rm s})}(x)\, .
\ea
In Appendix~\ref{app:analytic-continuation} we show that analytic continuation along the loop that goes around the singularity at $x=1$ counterclockwise gives
 \ba \label{Psi-monodromy-SQED}
\Psi^{({\rm s})}_\l(x) \r  \Psi^{({\rm s})}_\l(x)-
  \oint_{C'_2}d\eta
  \Bigg(
  \prod_{f=1}^{N_{\rm F}}\f{e^{2\pi i\eta}-e^{2\pi i\ti{m}_f}}{e^{2\pi i\eta}-e^{2\pi im_f}}
  \Bigg)
  \Psi^{({\rm s})}_{\eta}(x)\, ,
\ea
where the contour $C'_2$ encloses the poles $\{m_f | f =1,\ldots,N_{\rm F}\}$.

The relation (\ref{relesu2}) and the monodromy  (\ref{Psi-monodromy-SQED}) are equivalent to those obtained in \cite{Gomis:2014eya}. 
The vortex partition functions $f^{({\rm s})}_g(x)$ and $f^{({\rm u})}_g(x)$ that appear in the Higgs branch expression of the sphere partition of \cite{Gomis:2014eya} are proportional to specializations of the functions $\Psi_{\l}^{({\rm s})}(x)$ and $\Psi_{\xi}^{({\rm u})}(x)$ above to $\lambda=m_g$ and $\xi=\tilde{m}_g$ respectively.
In  Appendix \ref{rele-SQED} we explicitly confirm that the relations in (\ref{relesu2}) are equivalent to the relations between $f^{({\rm s})}_g(x)$ and $f^{({\rm u})}_g(x)$ given by the braiding matrices $B^{\pm}$ in \cite{Gomis:2014eya}, and that the monodromy matrix $M_1$ around $x=1$ as given in~\cite{Gomis:2014eya} is equivalent to our expression (\ref{Psi-monodromy-SQED}). 

\section{Janus interface and the generating function of A-model correlators}\label{sec:A-twist}

In this section we briefly study the Janus interface on $S^2$ with the equivariant A-twist, also known as the A-twist with omega deformation~\cite{Benini:2015noa,Closset:2015rna}.
We specialize to the Calabi-Yau hypersurface model considered in Section~\ref{sec:CY-hypersurface}.

The supergravity background, written down in~\cite{Closset:2014pda} and given in~\cite{Okuda:2017rwo} in the current convention, is characterized by the omega deformation parameter $\boldsymbol{\epsilon}_\Omega$.%
\footnote{%
The parameter $\boldsymbol{\epsilon}_\Omega$ was called $a$ in \cite{Okuda:2017rwo}.
}
(The special case $\boldsymbol{\epsilon}_\Omega=0$ corresponds to the pure A-twisted theory~\cite{Witten:1988xj,Morrison:1994fr}.)

In the absence of a Janus interface, the complexified FI parameter $t=r-i\vartheta$ is constant across the two-sphere.  Even in such a case, in a supersymmetric saddle point configuration, the scalar $\sigma$ in the vector multiplet is not constant when $\boldsymbol{\epsilon}_\Omega \neq 0 $.
Although SUSY localization is possible with the insertion of $\sigma$ either at the north $(\theta=0)$ or south ($\theta=\pi)$ pole, the value of the correlator does depend on at which pole $\sigma$ is inserted.
Let $\langle \mathcal{O} \rangle_t$ denote the expectation value of the operator $\mathcal{O}$ in the theory with modulus $t$.
The reference~\cite{Closset:2014pda} introduced the generating function
\begin{equation}\label{generating-function}
F(z;t) := \langle e^{z\sigma_{\rm N}} \rangle_t
= \sum_{n=0}^\infty \frac{z^n}{n!} \langle \sigma_{\rm N}^n \rangle_t
\end{equation}
of correlation functions in the theory, where $\sigma_{\rm N}$ is the field $\sigma$ inserted at the north pole.
It was found that $F_t(z)$ obeys an appropriate Picard-Fuchs equation.
These correlation functions were further studied in~\cite{Ueda:2016wfa,2016arXiv160708317K,Gerhardus:2018zwb}.

We point out that the generating function~(\ref{generating-function}) is nothing but the partition function in the presence of a Janus interface in the theory.

Let us construct the Janus interface in this background by the off-shell method we introduced in Section~\ref{sec:off-shell-construction}.
We promote $t$ to the scalar component of a twisted chiral multiplet $T$ and demand that the SUSY variations of the fermionic components vanish.
The trick we used in~(\ref{on-shell-tilde-W-exact}) and reviewed in footnote \ref{footnote:action-exact} gives the classical action in a supersymmetric configuration (saddle point in SUSY localization)
\begin{equation}
S_\text{Janus} = \frac{4\pi}{\boldsymbol{\epsilon}_\Omega} \widetilde{W}_\text{Janus} \Big|^{\theta=\pi}_{\theta=0} =
 \frac{1}{\boldsymbol{\epsilon}_\Omega}(t_{\rm N} \sigma_{\rm N} - t_{\rm S} \sigma_{\rm S}) \,,
\end{equation}
where the subscripts indicate that the symbols are evaluated at the north or the south pole.
We choose to rewrite this as
\begin{equation}
S_\text{Janus}  =
 \frac{t_{\rm N}-t_{\rm S}}{\boldsymbol{\epsilon}_\Omega}  \sigma_{\rm N} 
 +
  \frac{1}{\boldsymbol{\epsilon}_\Omega} t_{\rm S} (\sigma_{\rm N} -\sigma_{\rm S}) 
\end{equation}
and identify $z$ with $ (t_{\rm S}-t_{\rm N})/\boldsymbol{\epsilon}_\Omega $.
Then the Janus partition function with equivariant A-twist is
\begin{equation}
 Z_\text{Janus}^\text{A-twist, $\Omega$} = \langle e^{z\sigma_{\rm N}} \rangle_{t_{\rm S}} =  F(z;t_{\rm S})  \,.
\end{equation}

We also note that for the quintic model studied in~\cite{Closset:2015rna} the Picard-Fuchs equations are satisfied with respect to both $t_{\rm N}$ and $t_{\rm S}$.
Together with the results for correlators in~\cite{Closset:2015rna}, this implies that the Janus partition function can also be written as
\begin{equation} \label{Janus-A-twist-mirror}
Z_\text{Janus}^\text{A-twist, $\Omega$}  = \int \Omega(t_{\rm N})\wedge \Omega(t_{\rm S}) 
= X^I(t_{\rm N}) \mathcal{F}_I(t_{\rm S}) -  X^I(t_{\rm S}) \mathcal{F}_I(t_{\rm N})  \,,
\end{equation}
where $\Omega$ is the homomorphic three-form on the mirror Calabi-Yau such that $e^{-K} = i \int \Omega \wedge \overline{\Omega} $, and the periods $X^I$ and $\mathcal{F}_I$ are the quantities discussed in Section~\ref{sec:analytic-continuation}.
The relation between the generating function $F(z;t)$ (equivalent to an expression involving the Givental $I$-function or $\Psi_\lambda$) and the right-hand side of~(\ref{Janus-A-twist-mirror}) was observed earlier in (8.5.6) of~\cite{2016arXiv160708317K}, which proved the relation in the more general case.
We also note that the expression in the middle of (\ref{Janus-A-twist-mirror}) appeared already in the original paper~\cite{Candelas:1990rm}.

\section{Discussion}\label{sec:discussion}

As mentioned in the introduction, the localization result~(\ref{Z-Janus-Z-sphere}) and the value of the counterterm~(\ref{finite-counterterm-evaluated}) depend only on the values of $t$ and $\overline{t}$ at the north and south poles.
In a more refined treatment we expect the partition function should depend on the homotopy class of the path in the moduli space.
 
Let us review the argument for why the GLSM partition function computed by localization should coincide with the CFT sphere partition function.
The GLSM comes with the physical gauge coupling $g$ which has the dimension of mass.
If the size of the sphere is large compared with $g^{-1}$, the partition function of the GLSM should equal the CFT partition function.
In SUSY localization a new term ${\rm t} Q\cdot V$ with a fictitious coupling ${\rm t}$ is added to the action, so that the Gaussian approximation is exact in the limit ${\rm t}\rightarrow +\infty$.
Even though in this limit the characteristic length scale set by ${\rm t}$ becomes infinite, the modification of the action should not affect the value of path integral because it is $Q$-exact.

The length scale set by the variation of the moduli for the Janus interface is simply the radius of the sphere.
It would be interesting to pinpoint where the argument for the sphere partition function above fails for the Janus partition function with $t_{\rm N}=t_{\rm S}$ but with the profile $t(\theta)$ that gives a homotopically non-trivial path $\gamma$ in the moduli space.
We emphasize that we proposed and implemented a concrete mathematical analytic continuation procedure that computes the effect of monodromy.

A localization technique~\cite{Fourier-Mukai} that does capture the effect of monodromy is the introduction, in the folded theory, of a boundary interaction~\cite{Herbst:2008jq} based on the matrix factorization for the Fourier-Mukai transform corresponding to the monodromy~\cite{Brunner:2008fa}.
We can also consider a situation where the change in couplings occurs in several steps.
As in~\cite{Dedushenko:2018tgx} we may thicken the equator~$S^1$ into a thin strip $S^1\times [\frac{\pi}{2}-\epsilon,\frac{\pi}{2}+\epsilon]$ and integrate over suitable ``boundary conditions'' on the two boundaries $S^1\times \{\frac{\pi}{2}-\epsilon\}$ and $S^1\times \{\frac{\pi}{2}+\epsilon\}$ of the strip.
It is tempting to speculate that by letting $t(\theta)$ follow the path $\gamma$ as $\theta$ varies from $\frac{\pi}{2}-\epsilon$ to $\frac{\pi}{2}+\epsilon$, the degrees of freedom on the strip might yield the appropriate boundary interactions~\cite{Brunner:2008fa} that implement the Fourier-Mukai transform.
We leave the exploration of such possibilities to the future.

\section*{Acknowledgements}

The research of KG is supported in part by the JSPS Research Fellowship for Young Scientists.
The research of TO is supported in part by the JSPS Grants-in-Aid for Scientific Research No. 16K05312. 
TO thanks C.~Bachas, C.~Closset, E.~D'Hoker, B.~Le Floch, and Y.~Yoshida for useful discussion and correspondence.
We also thank B.~Le Floch for comments on an earlier draft.
TO acknowledges the Galileo Galilei Institute for Theoretical Physics for the hospitality and the INFN for partial support while this work was in progress.
 \appendix

\section{$\mathcal{N}=(2,2)$ supersymmetry and supergravity}

\subsection{Poincar\'e SUSY in Minkowski space}\label{sec:SUSY-Minkowski}

In Minkowski space we can write equations concisely in terms of superfields, which are functions of bosonic coordinates $x^\pm = x^0 \pm x^1$ and fermionic coordinates $(\theta^\pm, \overline{\theta}{}^\pm)$.
Supersymmetry transformations correspond to differential operators:
\begin{equation}
 Q_\pm \leftrightarrow  \frac{\partial}{\partial \theta^\pm} + i \overline{\theta}{}^\pm \partial_\pm \,,
 \qquad
 \overline{Q}_\pm \leftrightarrow - \frac{\partial}{\partial \overline{\theta}{}^\pm} - i \theta{}^\pm \partial_\pm \,.
\end{equation}
Various superfields are characterized by conditions involving superderivatives
\begin{equation}
 D_\pm =  \frac{\partial}{\partial \theta^\pm} - i \overline{\theta}{}^\pm \partial_\pm \,,
 \qquad
 \overline{D}_\pm = - \frac{\partial}{\partial \overline{\theta}{}^\pm} + i \theta{}^\pm \partial_\pm \,.
\end{equation}

In this paper twisted chiral superfields play prominent roles.
A twisted chiral superfield~$\widetilde{\Phi}$ is characterized by the conditions $\overline{D}{}_+\widetilde{\Phi} = D_- \widetilde{\Phi} =0 $.
It has an expansion
\begin{equation}
\begin{aligned}
\widetilde{\Phi}=v(\widetilde{y}) + \theta^+ \overline{\chi}{}_+(\widetilde{y})  +\overline{\theta}{}^- \chi{}_-(\widetilde{y}) +\theta^+\overline{\theta}{}^- E{}(\widetilde{y})  \,,
\end{aligned}
\end{equation}
where $\widetilde{y}^\pm = x^\pm \mp i\theta^\pm \overline{\theta}{}^\pm$.
A twisted superpotential $\widetilde{W}(\widetilde{\Phi})$ is a holomorphic function of twisted chiral superfields $\widetilde{\Phi}{}^j$.
The corresponding action in Minkowski space is
\begin{equation} \label{twisted-chiral-superpotential-Minkowski}
\begin{aligned}
S_{\widetilde{W}}&=
\int d^2x d^2\widetilde{\theta} \, \widetilde{W}(\widetilde{\Phi})+c.c. \\
 &=
 \int d^2x (E^j\pp_j\widetilde{W}+\chi^i_-\overline{\chi}{}_+^j\pp_i\pp_j\widetilde{W}+\overline{E}{}^j\pp_j\overline{\widetilde{W}}-\overline{\chi}{}^i_-\chi_+^j\pp_i\pp_j\overline{\widetilde{W}}) \,.
\end{aligned}
\end{equation}
In the last line the arguments of $\widetilde{W}$ and $\overline{\widetilde{W}}$ are $v^j$ and $\overline{v}{}^j$, respectively.
In our convention~$d^2\widetilde{\theta} =d\overline{\theta}{}^- d\theta^+$.

\subsection{Gauged linear sigma model}\label{sec:GLSM-review}

We review the gauged linear sigma model with the abelian gauge group~$U(1)^s$.
The formulas here are for Minkowski space.

The theory contains vector superfields $V_a$ ($a=1,\ldots,s$).
We can form twisted chiral superfields $\SG_a=\overline{D}_+D_-V_a$, which satisfiy the conditions $\overline{D}_+\SG_a=D_-\SG_a=0$. 
They can be expanded as
\begin{equation}
\SG_a
=\sg_a(\widetilde y)+i\theta^+\overline{\l}_{a+}( \widetilde y)-i \overline{\theta}{}^-\l_{a-}(\widetilde y)+\theta^+\overline{\theta}^-(D_a-iv_{a 01})(\widetilde y)
\,,
\end{equation}
where $v_{a\mu\nu} = \partial_\mu v_{a\nu} - \partial_\nu v_{a\mu}$ is the field stremgth for the gauge field $v_a=v_{a\mu}dx^\mu$.
Let $e_a$ be the gauge coupling for the $a$-th $U(1)$.  
The gauge kinetic term is
\begin{equation}
S_{\rm g}=\sum_{a}\frac{1}{2e^2_a}\int d^2x d^4\theta \overline{\SG}_a\SG_a \,.
\end{equation}
We also have the action~(\ref{twisted-chiral-superpotential-Minkowski}) constructed from the twisted superpotential
\begin{equation}
\widetilde{W}= - \frac{1}{4\pi} \sum_a  t^a\SG_a,\q t^a=r^a-i\vartheta^a \,.
\end{equation}
It is the sum of FI and theta terms
\begin{equation}
S_\text{FI-$\vartheta$}
=\int d^2x  \sum_a(-r^a D_a+\vartheta^a v_{a 01}) \,.
\end{equation}

The chiral superfields $\Phi^i$ are characterized by the conditions $\overline{D}_\pm \Phi^i=0$.
We denote their charges under the $a$-th $U(1)$ as $Q_{ia}$.  The chiral superfields can be expanded as 
\begin{equation}
\Phi^i=\phi^i(y)+\theta^\a\psi^i_{\a}(y)+ \theta^+\theta^-    F^i(y) \,,
\end{equation}
where $y^\pm = x^\pm - i \theta^\pm \overline{\theta}{}^\pm$.
The kinetic term for the chiral superfields is
\begin{equation}
S_{\rm m}=\sum_{i}\int d^2x d^4\theta \, \overline{\Phi}{}^ie^{Q_{ia} V_a}\Phi^i\,,
\end{equation}
where the fermionic measure is defined as $d^4\theta = d\theta^+ d\theta^- d\overline\theta{}^- d\overline\theta{}^+$.

A superpotential $W(\Phi)$ is a gauge invariant holomorphic function of $\Phi= (\Phi^i)$.
The corresponding action is
\begin{equation}
  S_W =  
  \int d^2x d^2 \theta  \, W(\Phi) + c.c. \,,
  \end{equation}
where $d^2\theta = d\theta^- d\theta^+$.

The total action for the gauged linear sigma model is
\begin{equation}
S= S_{\rm g}+S_{\rm m}+S_\text{FI-$\vartheta$}+S_W.
\end{equation}
 
\subsection{$\mathcal{N}=(2,2)$ $u(1)_\text{V}$ supergravity}

We collect relevant formulas for $\mathcal{N}=(2,2)$ $u(1)_\text{V}$ supergravity~\cite{Howe:1987ba,Grisaru:1994dm,Grisaru:1995dr,Gates:1995du,Ketov:1996es,Closset:2014pda}.
We use the convention of~\cite{Okuda:2017rwo}, and the formulas are given in Euclidean signature.
We set gravitini to zero as in~\cite{Closset:2014pda}.
We also assume that the SUSY parameters $(\epsilon_\mp, \overline{\epsilon}_\mp)$ are such that the gravitino variations vanish.
Our formulas are expressed in terms of the orthonormal frame $(e^{\hat z}, e^{\hat{\overline{z}}}) = (\Omega dz,\Omega d\overline{z})$ such that the metric is $ds^2 = e^{\hat z} e^{\hat{\overline{z}}} = \Omega^2 |dz|^2$.
Fields in twisted chiral and anti-chiral multiplets transform as
\begin{equation} \label{SUSY-twisted-chiral}
  \begin{aligned}
&
 \qquad \qquad \qquad
\delta v = \overline{\epsilon}_+ \chi_- -\epsilon_- \overline{\chi}_+\,, \\
&\delta\chi_- = 2i \epsilon_+ D_{\hat{z}} v +\epsilon_- E\,,
\qquad
\delta \overline{\chi}_+ 
= 2i \overline{\epsilon}_- D_{\hat{\overline{z}}}v + \overline{\epsilon}_+ E\,,\\
&
\qquad \qquad
\delta E = 2i \epsilon_+ D_{\hat z} \overline{\chi}_+ -2i \overline{\epsilon}_- D_{\hat{\overline{z}}} \chi_-\, ,
  \end{aligned}
\end{equation}
\begin{equation}  \label{SUSY-twisted-anti-chiral}
  \begin{aligned}
&\ \  \qquad\qquad \qquad  \delta \overline{v}  = - \epsilon_+ \overline{\chi}_- + \overline{\epsilon}_- \chi_+ \,, \\
&  \delta \overline{\chi}_-  = - 2i \overline{\epsilon}_+ D_{\hat{z}} \overline{v} + \overline{\epsilon}_- \overline{E} \,, 
\qquad
  \delta \chi_+  = - 2i \epsilon_- D_{\hat{\overline{z}}} \overline{v} + \epsilon_+ \overline{E} \,, \\
&\qquad\qquad \qquad  \delta \overline{E}  = 2i \overline{\epsilon}_+ D_{\hat{z}} \chi_+ - 2i \epsilon_- D_{\hat{\overline{z}}} \overline{\chi}_- \,.
  \end{aligned}
  \end{equation}
 
 Given a twisted superpotential $\widetilde{W}(v)$ for twisted chiral multiplets $(v^j, \chi^j_-, \overline{\chi}^j_+, E^j)$, the corresponding action in Euclidean signature is \begin{equation} \label{twisted-F-term-coupling-rigid}
S_{\widetilde{W}} = -  \int d^2x \sqrt{g}\Big( E^j\partial_j \widetilde{W}  + \chi^i_- \overline{\chi}^j_+ \partial_i \partial_j \widetilde{W}    + \overline{\mathcal{H}}\widetilde{W} + \overline{E}^j \partial_j \overline{\widetilde{W}}   - \overline{\chi}^i_- \chi_+^j \partial_i \partial_j \overline{\widetilde{W}}   + \mathcal{H} \overline{\widetilde{W}} \Big)  \,,
\end{equation}
where $\mathcal{H}$ is the graviphoton field strength.
On the deformed sphere~(\ref{deformed-sphere-metric}), $\mathcal{H}$ and its conjugate take value $\mathcal{H}=\overline{\mathcal{H}}=-i/{\rm f}(\theta)$.
 
 \section{Details of analytic continuation}\label{app:analytic-continuation}
 
In this appendix we provide details of the analytic continuation of $\Psi_\lambda(x)$.
Our analysis is most directly influenced by~\cite{Horja-hypergeometric}.
The formulas for analytic continuation and monodromies in this appendix have been checked by plotting the functions along continuation paths numerically.

\subsection{GLSM for the Calabi-Yau hypersurface}

Recall from (\ref{Psi-lambda-def}) that 
\begin{equation} \label{Psi-lambda-def-app}
\Psi_\lambda(x) = \sum_{k=0}^\infty \frac{\Gamma(1+ n(k+\lambda))}{\Gamma(1+k+\lambda)^n} x^{k+\lambda} 
\end{equation}
in the region near the large volume point $x=0$.
The function has a singularity at $x=n^{-n}$.

For the region near the Landau-Ginzburg point $x=\infty$, we define
\begin{equation} \label{Psi-LG-def}
\Psi^\text{LG}_\xi(x):= \sum_{k=0}^\infty \frac{\Gamma(k  - \xi)^n}{\Gamma(n(k  - \xi))} x^{-k+\xi} \,.
\end{equation}

Let us assume that $0< {\rm arg}\,x < 2\pi$.
We can write
\begin{equation}
\Psi_\lambda(x) =  \int \frac{ds}{2\pi i} \frac{\Gamma( n(s+\lambda) +1 )}{\Gamma(s+\lambda+1)^n } x^{s+\lambda} e^{-\pi i s} \Gamma(-s)\Gamma(1+s) \,.
\end{equation}
The contour asymptotes to the imaginary axis and is chosen to separate the poles of $\Gamma(-s)$ from those of $\Gamma( n(s+\lambda) +1 )$ and~$\Gamma(1+s)$.
The integral is absolutely convergent for $0< {\rm arg}\,x < 2\pi$.
For $|x|< n^{-n}$ we can close the contour to the right and reproduce the series definition (\ref{Psi-lambda-def-app}).

For $|x| > n^{-n}$  we can close the contour to the left and pick up the poles of $\Gamma( n(s+\lambda) +1 )$ and~$\Gamma(1+s)$.
There are $n$ sequences of poles extending to the left with integer steps, starting at an element of $\{-\frac{1}{n}, \ldots, - \frac{n-1}{n}, \lambda -1 \}$.
We can write the result using (\ref{Psi-LG-def}) as
\begin{equation} \label{Psi-PsiLG-3}
\Psi_\lambda(x) = \oint_{C_1} d\xi \frac{(1-e^{2\pi i \xi})^n}{(2\pi i)^{n-1}(1- e^{ 2 n \pi i \xi}) ( 1- e^{ 2\pi i (\xi - \lambda)})}
\Psi^\text{LG}_\xi(x) \,.
\end{equation}
The contour $C_1$ encircles counterclockwise the poles $\{-\frac{1}{n}, \ldots, - \frac{n-1}{n}, \lambda -1 \}$ but no other poles.
Note in particular that $\Psi^\text{LG}_\xi(x)$ as a function of $\xi$ has singularities at $\xi \in  \mathbb{Z}_{\geq 0}$; the contour $C_1$ should not enclose them.

A similar consideration in the opposite direction gives
\begin{equation}\label{PsiLG-Psi-3}
\Psi^\text{LG}_\xi(x) = \oint_{C_2} d\mu
\frac{ (2\pi i)^{n-1} (1-e^{2n\pi i \mu})}{(1-e^{2\pi i \mu})^n(1 - e^{2\pi i (\mu-\xi)})}
\Psi_\mu(x) \,.
\end{equation}
The contour $C_2$ encloses $0$ and $\xi+1$ counterclockwise.
As a function of $\mu$, $\Psi_\mu(z^{-1})$ has singularities at $\mu \in -\frac{1}{n}  \mathbb{Z}_{\geq 1} \backslash - \mathbb{Z}_{\geq 1}$, which the contour $C_2$ should not enclose.

\subsubsection{Trivial loop}

Let us consider a homotopically trivial contour in the $x$-plane that starts at a point near the origin, goes out to a region with a large absolute value, and then comes back to the original point, always with the argument in the range $0< {\rm arg}\,x < 2\pi$.
This corresponds to the integral
\begin{equation}
I= \oint_{C_1} d\xi 
  \oint_{C_2} d\mu
 \frac{(1-e^{2\pi i \xi})^n}{(1- e^{ 2 n \pi i \xi}) ( 1- e^{ 2\pi i (\xi - \lambda)})}
 \frac{ (1-e^{2n\pi i \mu})}{(1-e^{2\pi i \mu})^n(1 - e^{2\pi i (\mu-\xi)})}
 \Psi_\mu(x) \,.
\end{equation}
We deform $C_2$ to a new contour $C'_2$ so that it encloses the contour $C_1$ and its shifted copy $C_1+1$.
Upon deformation we pick up the residue of the pole at $\mu=\xi$, which contributes upon $\xi$-integration
\begin{equation}
 \oint_{C_1} d\xi 
 \frac{-1}{ ( 1- e^{ 2\pi i (\xi - \lambda)})}
 \Psi_\xi(x) 
 =   \Psi_{\lambda-1}(x) \,,
\end{equation}
which we need to subtract.
We have
\begin{equation}
\begin{aligned}
I &= 
   \oint_{C_1} d\xi 
     \oint_{C'_2} d\mu
 \frac{(1-e^{2\pi i \xi})^n}{(2\pi i)^{n-1}(1- e^{ 2 n \pi i \xi}) ( 1- e^{ 2\pi i (\xi - \lambda)})}
 \\
 &\qquad\qquad\qquad\qquad
 \times
 \frac{ (2\pi i)^{n-1} (1-e^{2n\pi i \mu})}{(1-e^{2\pi i \mu})^n(1 - e^{2\pi i (\mu-\xi)})}
 \Psi_\mu(x)
- \Psi_{\lambda-1}(x) \,.
\end{aligned}
\end{equation}
We can now change the order of integrations.
We also change a variable from $\xi$ to $u=e^{2\pi i \xi}$.
We get
\begin{equation}
\begin{aligned}
I&=
  \oint_{C'_2} d\mu
   \oint \frac{du}{2\pi i u}
 \frac{(1-u)^n}{(2\pi i)^{n-1}(1- u^n) ( 1-  u e^{ - 2\pi i \lambda})}
 \\
 &\qquad\qquad\qquad\qquad\times
 \frac{ (2\pi i)^{n-1} (1-e^{2n\pi i \mu})}{(1-e^{2\pi i \mu})^n(1 - e^{2\pi i \mu}u^{-1})}
 \Psi_\mu(x) 
 - \Psi_{\lambda-1}(x)
   \,.
   \end{aligned}
\end{equation}
The $u$-contour encloses $e^{- \frac{2\pi i}{n} j}$, $j=1,\ldots,n-1$, and $e^{2\pi i \lambda}$.
We want to pull it out and evaluate the $u$-integral in terms of the poles outside the $u$-contour.
There is no pole at $u=\infty$ or at $u=1$.
Only the pole at $u=e^{2\pi i \mu}$ contributes.
We get
\begin{equation}
I =   \oint_{C'_2} d\mu
 \frac{1}{   e^{  2\pi i (\mu - \lambda)}- 1}
 \Psi_\mu(x) 
 - \Psi_{\lambda-1}(x)
 \,.
\end{equation}
Since $C_1$ enclosed $\lambda-1$, $C'_2$ encloses $\lambda-1$ and $\lambda$.
Thus we obtain
\begin{equation}
I =\Psi_\lambda(x)  
\end{equation}
as expected for analytic continuation along a homotopically trivial contour.

\subsubsection{Loop around the conifold singularity}

Next, let us consider analytically continuing $x$ first around the origin clockwise, and then around a large circle counterclockwise.

\begin{figure}[htbp]
\centering
\includegraphics[width=6cm]{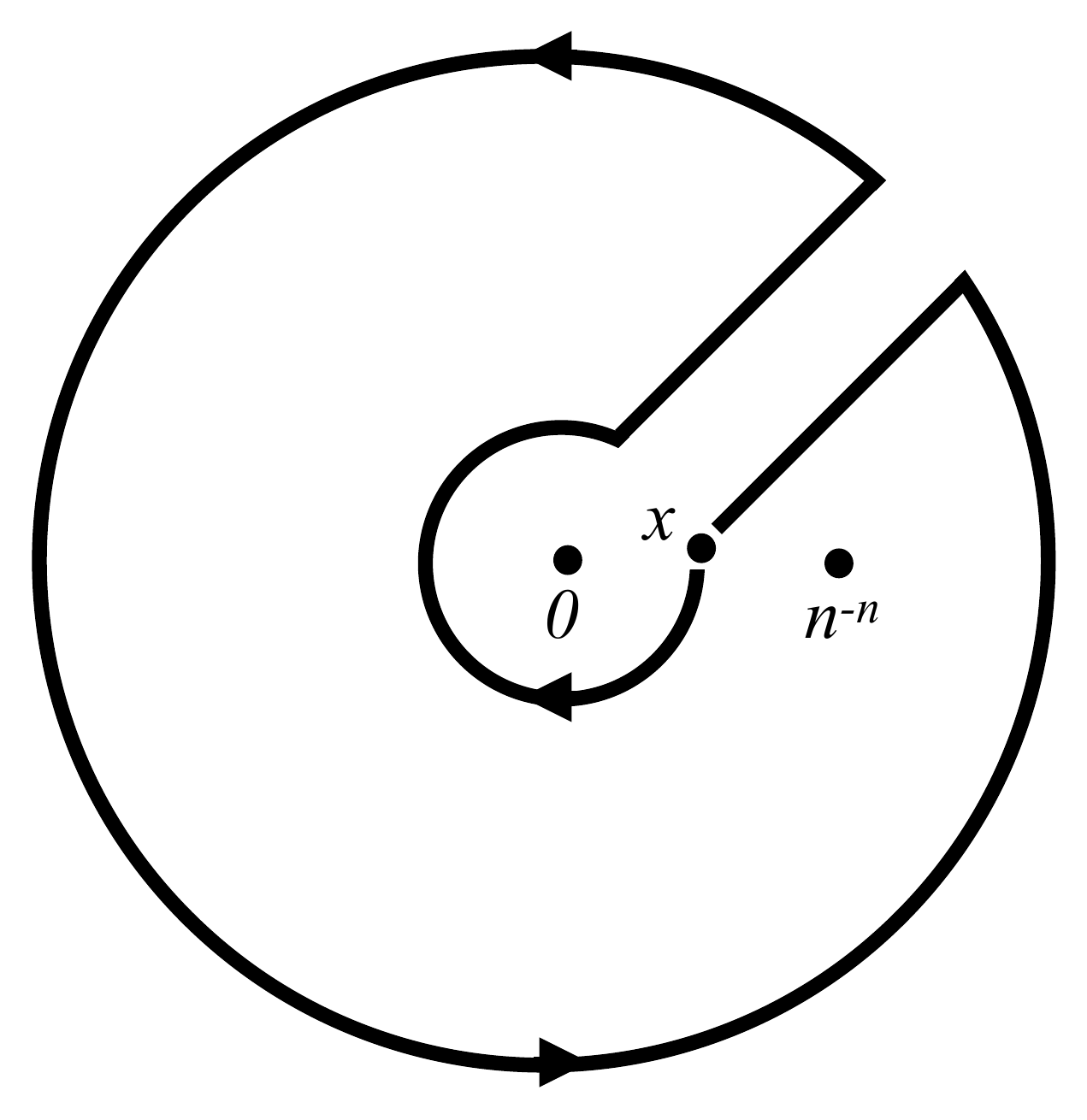}
\caption{\label{figure:conifold-loop-x}%
Analytic continuation path for the non-trivial loop counterclockwise around the singularity at $x=n^{-n}$.
}
\end{figure}

We begin with a point near the origin with $0< {\rm arg}\,x < 2\pi$.
Continuing around the origin clockwise, we get
\begin{equation}
\Psi_\lambda(x) = e^{-2\pi i\lambda}\Psi_\lambda(e^{2\pi i} x) \,.
\end{equation}
Note that $0< {\rm arg}(e^{2\pi i} x) <2\pi$ at this stage.
Moving outward with fixed ${\rm arg} \, x$, we get for $|x|> n^{-n}$
\begin{equation}
\Psi_\lambda(e^{2\pi i} x) 
=
 \oint_{C_1} d\xi \frac{(1-e^{2\pi i \xi})^n}{(2\pi i)^{n-1}(1- e^{ 2 n \pi i \xi}) ( 1- e^{ 2\pi i (\xi - \lambda)})}
 \Psi^\text{LG}_\xi(e^{2\pi i}x) \,.
\end{equation}
We then continue along a large circle counterclockwise using
\begin{equation}
 \Psi^\text{LG}_\xi(e^{2\pi i}x) 
  =e^{2\pi i \xi}
 \Psi^\text{LG}_\xi(x)  \,.
\end{equation}
At this point we have $0< {\rm arg}\, x<2\pi$.
We then move inward, keeping ${\rm arg}\, x$ fixed, to the region $|x|< n^{-n}$:
\begin{equation}
\Psi^\text{LG}_\xi( x) = \oint_{C_2} d\mu
\frac{ (2\pi i)^{n-1} (1-e^{2n\pi i \mu})}{(1-e^{2\pi i \mu})^n(1 - e^{2\pi i (\mu-\xi)})}
\Psi_\mu(x) \,.
\end{equation}
Thus the result of analytic continuation is
\begin{equation} \label{eq:Psi-conifold-continuation}
\begin{aligned}
\Psi_\lambda(x)
&
\rightarrow
e^{-2\pi i\lambda}
 \oint_{C_1} d\xi \frac{(1-e^{2\pi i \xi})^n}{(2\pi i)^{n-1}(1- e^{ 2 n \pi i \xi}) ( 1- e^{ 2\pi i (\xi - \lambda)})}
\\
&
\qquad\qquad\qquad\times
e^{2\pi i \xi} \oint_{C_2} d\mu
\frac{ (2\pi i)^{n-1} (1-e^{2n\pi i \mu})}{(1-e^{2\pi i \mu})^n(1 - e^{2\pi i (\mu-\xi)})}
\Psi_\mu(x) \,.
\end{aligned}
\end{equation}
We evaluate the right-hand side as before, by deforming $C_2$ to $C'_2$ and by changing the order of integrations.
We get
\begin{equation}
\begin{aligned}
\text{RHS of (\ref{eq:Psi-conifold-continuation})}
&=
e^{-2\pi i\lambda}
 \oint_{C'_2} d\mu
  \oint
  \frac{du}{2\pi i u}
  \frac{(1-u)^n}{(2\pi i)^{n-1}(1- u^n) ( 1-  u e^{-2\pi i   \lambda})}
    \\
 &\qquad\qquad\qquad\qquad\times
u
\frac{ (2\pi i)^{n-1} (1-e^{2n\pi i \mu})}{(1-e^{2\pi i \mu})^n(1 - e^{2\pi i \mu}u^{-1})}
\Psi_\mu(x)
-
\Psi_{\lambda-1}(x) \,.
\end{aligned}
\end{equation}
Here the $u$-contour
encloses  $e^{-2\pi i \frac{b}{n}}$ ($b=1,\ldots,n-1$) and $e^{2\pi i \lambda}$ counterclockwise.
Upon pulling it out, the pole at $u=\infty$ contributes
\begin{equation}
(-1)^n
 \oint_{C'_2} d\mu
\frac{  (1-e^{2n\pi i \mu})}{(1-e^{2\pi i \mu})^n}
\Psi_\mu(x) 
=
-
 \oint_{C'_2} d\mu
\frac{  1 - e^{-2n\pi i \mu}}{(1- e^{-2\pi i \mu})^n}
\Psi_\mu(x) 
\,.
\end{equation}
Though $C'_2$ encloses $0$, $\lambda-1$, and $\lambda$, we can deform it so that it encloses only $0$ because there is no pole at $\mu= \lambda-1$ or at $\mu=\lambda$.
The remaining part is calculated as before.
This leads to the analytic continuation formula (\ref{Psi-monodromy-conifold}).

\subsection{SQED}

Here we perform analytic continuation for SQED studied in Section~\ref{sec:SQED}.
The manipulations are similar to those for the Calabi-Yau hypersurface above.

\subsubsection{$\Psi_{\l}^{({\rm s})}(x)$ and $\Psi_{\zeta}^{({\rm u})}(x)$} \label{sec:SQED-Psi}

In (\ref{Zsp}) and  (\ref{Zup}) we defined the functions $\Psi_{\l}^{({\rm s})}(x)$ and $\Psi_{\zeta}^{({\rm u})}(x)$ as convergent series for $|x|<1$ and $|x|>1$, respectively.
Let us assume that $0<\arg{x}<2\pi$.
In this region the integral 
\begin{equation}
\int^{i\infty}_{-i\infty} \frac{ds}{2\pi i}
\Bigg(\prod_{f=1}^{N_{\rm F}}\f{\G(s+\l-\ti{m}_f)}{\G(s+1+\l-m_f)}\Bigg)\Gamma(-s)\Gamma(1+s)e^{-\pi i s} 
\end{equation}
absolutely converges.
For $|x|<1$ we can close the contour to the right and recover the series definition for $\Psi_{\l}^{({\rm s})}(x)$.
For $|x|>1$, closing the contour to the left leads to the relation between  $\Psi_{\l}^{({\rm s})}(x)$ and $\Psi_{\zeta}^{({\rm u})}(x)$ in the first line of~(\ref{relesu2}).
The relation in the second line can be derived similarly.

\subsubsection{Trivial loop}

\begin{figure}[htbp]
\centering
\includegraphics[width=13cm]{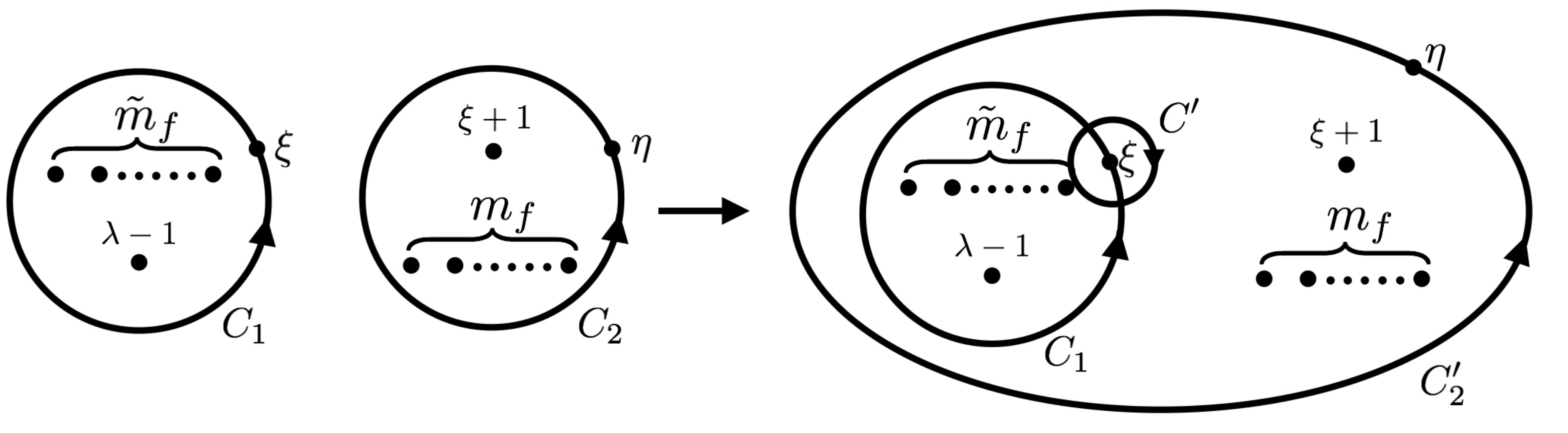}
\caption{\label{figure:contourCCC}%
Deformation of the integration contour $C_2$ to $C_2'$ and $C'$
}
\end{figure}

We first consider a homotopically trivial contour that corresponds to the integral
\begin{equation}
\begin{aligned}
 &
 I: =\oint_{C_1}\f{d\xi}{2\pi i}\oint_{C_2}\f{d\eta}{2\pi i}
  \Bigg(
 \prod_{f=1}^{N_{\rm F}}\f{\sin\pi(\xi-m_f)}{\sin\pi(\xi-\ti{m}_f)}\f{\sin\pi(\eta-\ti{m}_f)}{\sin\pi(\eta-m_f)}
 \Bigg)
 \\
 &\qquad\qquad \qquad\qquad \qquad\qquad \qquad\qquad 
 \times
 \f{\pi e^{-\pi i (\xi-\l)}}{\sin\pi (\xi-\l)}\f{\pi e^{- \pi i (\eta-\xi)}}{\sin\pi (\eta-\xi)}\Psi^{({\rm s})}_{\eta}(x)\, .
 \end{aligned}
\end{equation}
 Contour $C_1$ encloses $\{\ti{m}_f|f=1,\ldots,N_{\rm F}]\}\cup\{\l-1\}$ while $C_2$ encloses $\{m_f|f=1,\ldots,N_{\rm F}\}\cup\{\xi + 1\}$. 
See Figure~\ref{figure:contourCCC}.
 We deform $C_2$ to $C'_2$ so that it also encloses $C_1$ as in Figure~\ref{figure:contourCCC}. 
 Since $C_2=C'_2+C'$ homologically, we can write 
 \begin{equation}
 I= I_1 + I_2\,,
 \end{equation}
where
 \begin{equation}
 \begin{aligned}
I_1&:=
\oint_{C_2'}\f{d\eta}{2\pi i}\oint_{C_1}\f{d\xi}{2\pi i}
\Bigg(
\prod_{f=1}^{N_{\rm F}}\f{\sin\pi(\xi-m_f)}{\sin\pi(\xi-\ti{m}_f)}\f{\sin\pi(\eta-\ti{m}_f)}{\sin\pi(\eta-m_f)}
\Bigg)
\\
&\qquad\qquad\qquad\qquad\qquad\qquad\qquad
\times 
\f{\pi e^{-\pi i (\xi-\l)}}{\sin\pi (\xi-\l)}\f{\pi e^{- \pi i (\eta-\xi)}}{\sin\pi (\eta-\xi)}\Psi^{({\rm s})}_{\eta}(x)
 \end{aligned}
\end{equation}
and
\begin{equation}
I_2:=-\oint_{C_1}\f{d\xi}{2\pi i}\f{\pi e^{-\pi i (\xi-\l)}}{\sin\pi (\xi-\l)}\Psi^{({\rm s})}_{\xi}(x)\, .
\end{equation}
Note that for $I_1$ we have changed the order of integrations, and for $I_2$ we have performed the integration along $C'$ by picking up the residue at $\eta=\xi$ and there only remains a single integral with respect to $\xi$.

In terms of a new integration variable $u=e^{2\pi i \xi}$, we have
\begin{equation}
\begin{aligned}
I_1&=\oint_{C'_2}\f{d\eta}{2\pi i}\prod_{f=1}^{N_{\rm F}}\f{\sin\pi(\eta-\ti{m}_f)}{\sin\pi(\eta-m_f)}\Psi^{({\rm s})}_{\eta}(x)\\
&\qquad\qquad\times
\oint_{C_1}\f{du}{(2\pi i)^2u}
\Bigg(
\prod_{f=1}^{N_{\rm F}}\f{ue^{-\pi im_f}-e^{\pi im_f}}{ue^{-\pi i\ti{m}_f}-e^{\pi i\ti{m}_f}}
\Bigg)
 \f{4\pi^2 ue^{-\pi i (\eta-\l)}}{(e^{i\pi\l}-e^{-i\pi\l}u)(e^{i\pi\eta}-e^{-i\pi\eta}u)}\, .
\end{aligned}
\end{equation}
Outside $C_1$, $u=\infty$ is no longer a pole, and evaluating the residue at the pole $u=e^{2\pi i\eta}$ gives
\bal
I_1 
 =\oint_{C'_2}\f{d\eta}{2\pi i}\f{\pi e^{-\pi i (\eta-\l)}}{\sin\pi (\eta-\l)}\Psi^{({\rm s})}_{\eta}(x)=\Psi^{({\rm s})}_{\l}(x)+\Psi^{({\rm s})}_{\l-1}(x)\, .
\eal
The integral $I_2$ simply gives
$-\Psi^{({\rm s})}_{\l-1}(x)$. 
Combining $I_1$ and $I_2$, we get 
\ba
I=\Psi^{({\rm s})}_{\l}(x)
\ea
as expected for the trivial contour.
\subsubsection{Loop around the singularity at $x=1$}
Next, let us consider analytically continuing $x$ first around the origin clockwise, and then around a large circle counterclockwise as we did for the Calabi-Yau case.
\begin{figure}[htbp]
\centering
\includegraphics[width=6cm]{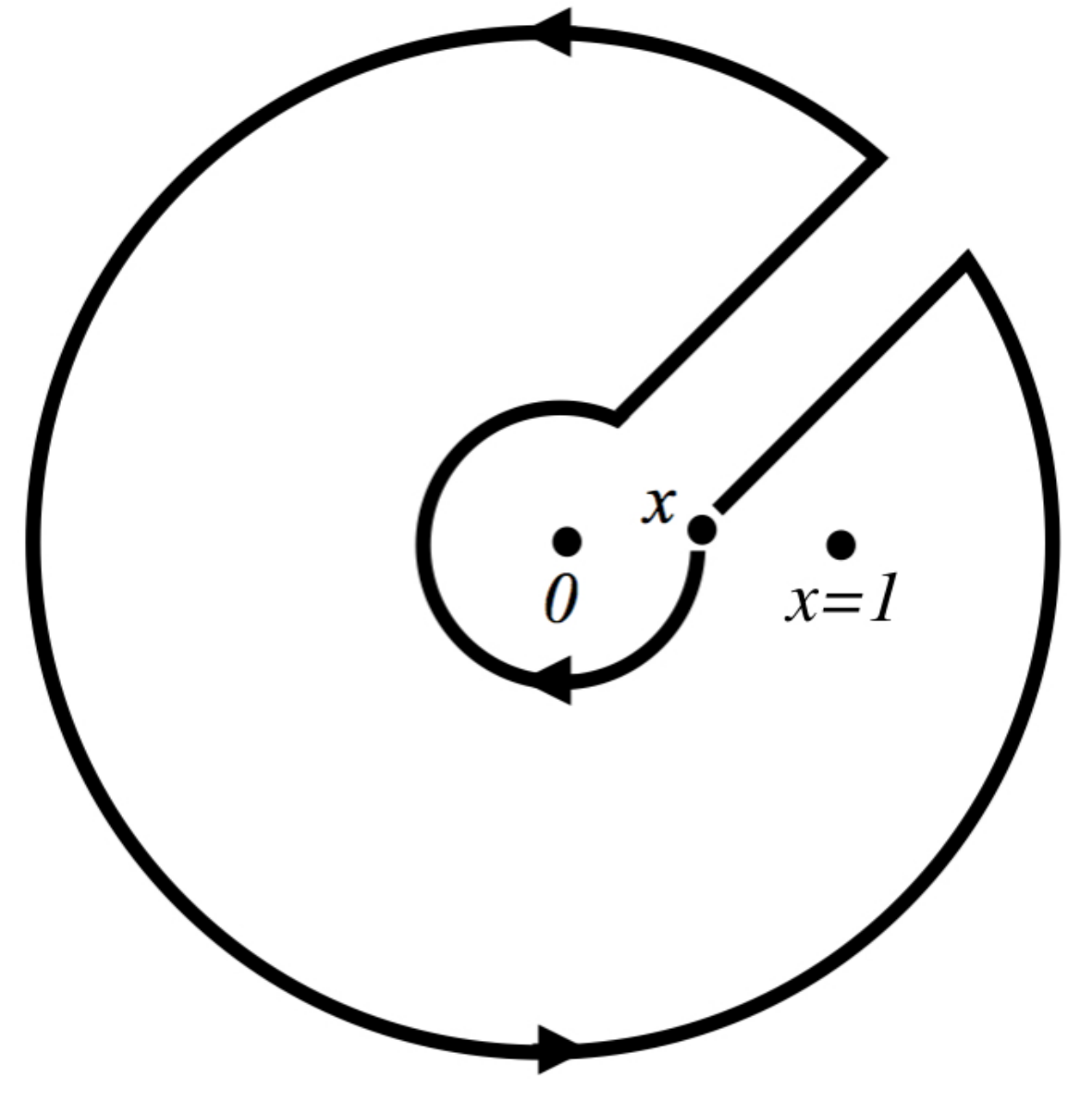}
\caption{\label{figure:loop-x=1}%
Analytic continuation path for the non-trivial loop counterclockwise around the singularity at $x=1$.
}
\end{figure}

We begin with a point near the origin with $0 < \arg{x} < 2\pi$. In the same way as the Calabi-Yau case, first by continuing around the origin clockwise we have
\ba
 \Psi^{({\rm s})}_\l(x) =e^{-2\pi i \l} \Psi^{({\rm s})}_\l(e^{2\pi i}x)\,.
\ea
Moving outward to the region $|x|>1$ with fixed $\arg{x}$, we use the first line of~(\ref{relesu2}).
Continuing along a large circle counterclockwise we have
 \ba
 \Psi^{({\rm u})}_\xi(e^{2\pi i}x) =e^{2\pi i \xi} \Psi^{({\rm u})}_\xi(x) \,.
\ea
Finally we use the second line of~(\ref{relesu2}) to move inward to the region $|x|<1$ with fixed $\arg{x}$.
Combining the operations we get
\begin{equation}
\begin{aligned}
\Psi^{({\rm s})}_\l(x) 
&\r e^{-2\pi i \l}\oint_{C_1}\f{d\xi}{2\pi i}
\Bigg(
\prod_{f=1}^{N_{\rm F}}\f{\sin\pi(\xi-m_f)}{\sin\pi(\xi-\ti{m}_f)}
\Bigg)
\f{(-1)^{N_{\rm F}}\pi e^{-\pi i (\xi-\l)}}{\sin\pi (\xi-\l)}
\\
&\qquad\qquad\qquad
\times e^{2\pi i \xi} \oint_{C_2}\f{d\eta}{2\pi i}
\Bigg(
\prod_{f=1}^{N_{\rm F}}\f{\sin\pi(\eta-\ti{m}_f)}{\sin\pi(\eta-m_f)}
\Bigg)
\f{(-1)^{N_{\rm F}}\pi e^{- \pi i (\eta-\xi)}}{\sin\pi (\eta-\xi)}\Psi^{({\rm s})}_{\eta}(x).
\end{aligned}
\end{equation}
We evaluate the right-hand side as we did for the trivial loop. 
In terms of
 $u=e^{2\pi i \xi}$, we have
\begin{equation}
\begin{aligned}
 \Psi^{({\rm s})}_\l(x) &\r \oint_{C'_2}\f{d\eta}{2\pi i}
 \Bigg(
 \prod_{f=1}^{N_{\rm F}}\f{\sin\pi(\eta-\ti{m}_f)}{\sin\pi(\eta-m_f)}
 \Bigg)
 \Psi^{({\rm s})}_{\eta}(x)
 \oint_{C_1}\f{du}{(2\pi i)^2u}
 \Bigg(
 \prod_{f=1}^{N_{\rm F}}\f{ue^{-\pi im_f}-e^{\pi im_f}}{ue^{-\pi i\ti{m}_f}-e^{\pi i\ti{m}_f}}
 \Bigg)
 \\
 &\qquad \qquad 
 \times
 \f{4\pi^2 u^2e^{-\pi i (\l+\eta)}}{(e^{i\pi\l}-e^{-i\pi\l}u)(e^{i\pi\eta}-e^{-i\pi\eta}u)} 
 - \oint_{C_1}\f{d\xi}{2\pi i}\f{\pi e^{\pi i (\xi-\l)}}{\sin\pi (\xi-\l)}\Psi^{({\rm s})}_{\xi}(x)\, .
\end{aligned}
\end{equation}
We deform $C_1$ to the contour which encloses poles at $u=e^{2\pi i\eta}$ and $u=\infty$. 
The combination of the contributions from the pole at $u=e^{2\pi i\eta}$ and the $\xi$-integral in the second line gives $\Psi^{({\rm s})}_\l(x)$ as we discussed for the trivial loop. It is convenient to introduce an integration variable $v=1/u$ for the integral around $u=\infty$. We can write the integral as
\begin{equation}
\begin{aligned}
\Psi^{({\rm s})}_\l(x) &\r  \Psi^{({\rm s})}_\l(x)+
  \oint_{C'_2}\f{d\eta}{2\pi i}
  \Bigg(
  \prod_{f=1}^{N_{\rm F}}\f{\sin\pi(\eta-\ti{m}_f)}{\sin\pi(\eta-m_f)}
  \Bigg)
  \Psi^{({\rm s})}_{\eta}(x)
  \\
  &
  \qquad
  \times\oint_{0}\f{dv}{(2\pi i)^2v}
\Bigg(
\prod_{f=1}^{N_{\rm F}}\f{e^{-\pi im_f}-e^{\pi im_f}v}{e^{-\pi i\ti{m}_f}-e^{\pi i\ti{m}_f}v}
\Bigg)
 \f{4\pi^2 e^{-\pi i (\l+\eta)}}{(e^{i\pi\l}v-e^{-i\pi\l})(e^{i\pi\eta}v-e^{-i\pi\eta})}\, .
 \end{aligned}
\end{equation}
By picking up the residue at $v=0$, we finally obtain the monodromy formula~(\ref{Psi-monodromy-SQED}).

\subsubsection{Relation to the braiding and monodromy matrices}\label{rele-SQED}

In this appendix, we will compute the matrices that represent the relations~(\ref{relesu2}) between $\Psi^{({\rm s})}_\l(x)$ and $\Psi^{({\rm u})}_\xi(x)$,
as well as the matrix representing the monodromy~(\ref{Psi-monodromy-SQED}).
We will compare them with the corresponding expressions found in Appendix A of \cite{Gomis:2014eya}.

Slightly generalizing the first line of~(\ref{relesu2}), we can show that
 \bal
   \Psi^{({\rm s})}_\l(x)&=-\oint_{C_1}\f{d\xi}{2\pi i}
\Bigg(
\prod_{f=1}^{N_{\rm F}}\f{\sin\pi(\xi-m_f)}{\sin\pi(\xi-\ti{m}_f)}
\Bigg)
\f{(-1)^{N_{\rm F}}\pi e^{ \mp \pi i (\xi-\l)}}{\sin\pi (\xi-\l)}\Psi^{({\rm u})}_{\xi}(x)\, ,
\eal
where $C_1$ encloses all the poles $\xi=\tilde{m}_f$, and the upper or lower sign corresponds to $0<\arg(x)<2\pi$ or $-2\pi <\arg(x)<0$ respectively. 
After performing integration we have
 \bal
   \Psi^{({\rm s})}_{m_g}(x)&=-\sum_{f=1}^{N_{\rm F}}\f{\prod_{h=1}^{N_{\rm F}}\sin\pi(\tilde{m}_f-m_h)}{\prod_{h\neq f}^{N_{\rm F}}\sin\pi(\tilde{m}_f-\ti{m}_h)}\f{(-1)^{N_{\rm F}} e^{\pm i\pi  (m_g-\tilde{m}_f)}}{\sin\pi (-\tilde{m}_f+m_g)}\Psi^{({\rm u})}_{\tilde{m}_f}(x)\, .\label{relesu}
\eal
Let us introduce the functions (vortex partition functions~\cite{Shadchin:2006yz})
\begin{equation}
\begin{aligned}
f^{({\rm s})}_g(x)&:=
F\left(\begin{matrix} m_g-\tilde{m}_{f}, 1\leq f\leq N_\text{F}\\ 1-m_{f}+m_g, f\neq p\end{matrix};x\right)
=x^{-m_g}\Psi^{({\rm s})}_{m_g}(x)\prod_{f=1}^{N_\text{F}}\f{\Gamma(1-m_f+m_g)}{\Gamma(-\tilde{m}_f+m_g)}\, ,\no
f^{({\rm u})}_g(x)&:=
  F\left(\begin{matrix} \tilde{m}_g-m_{f}, 1\leq s\leq N_\text{F}\\ 1+\tilde{m}_{f}-\tilde{m}_g, s\neq p\end{matrix};\f{1}{x}\right)
=x^{-\tilde{m}_g}\Psi^{({\rm u})}_{\tilde{m}_g}(x)\prod_{f=1}^{N_\text{F}}\f{\Gamma(1+\tilde{m}_f-\tilde{m}_g)}{\Gamma(m_f-\tilde{m}_g)}\,,
\end{aligned}
\end{equation}
where $F\left(
\begin{matrix} a_1\ a_2 \ddd a_{N_\text{F}} \\ b_1\ b_2 \ddd b_{N_\text{F}-1} \end{matrix}; x\right)$ denotes the generalized hypergeometric function.
Using these functions, we can rewrite (\ref{relesu}) as
\begin{equation} \label{f-s-u}
x^{m_g}f^{({\rm s})}_g(x)
=\sum_{f=1}^{N_\text{F}} B^\pm_{gf} \, x^{\tilde{m}_f}f^{({\rm u})}_f(x)\, ,
\end{equation}
where we defined
\begin{equation} \label{braiding-matrix-def}
\begin{aligned}
&
\ D_f:= \prod_{g=1}^{N_{\rm F}}\f{\Gamma(1-m_g+m_f)}{\Gamma(-\tilde{m}_g+m_f)}\,,
\qquad
\tilde{D}_f:=\f{\prod_{g\neq f}^{N_{\rm F}}\Gamma(-\tilde{m}_g+\tilde{m}_f)}{\prod_{g=1}^{N_{\rm F}}\Gamma(1-m_g+\tilde{m}_f)}  \\
&
\qquad\quad
\check{B}^{\pm}_{gf}:=\f{\pi e^{\pm i\pi  (m_g-\tilde{m}_f)}}{\sin\pi (-\tilde{m}_f+m_g)}
\,,
\qquad
B^\pm_{gf}:= D_g\check{B}^{\pm}_{gf}\tilde{D}_f\, .
\end{aligned}
\end{equation}
The definition~(\ref{braiding-matrix-def}) of the braiding matrix $B^\pm$ coincides with that in (A.3.3) of \cite{Gomis:2014eya}.

The relation inverse to~(\ref{f-s-u}) can be written as
\ba
x^{\tilde{m}_g}f^{({\rm u})}_g(x)= \sum_{f=1}^{N_\text{F}}  (B^\pm)^{-1}_{gf} x^{m_f}f^{({\rm s})}_f(x)\,.
\ea
The analog of the manipulations that led to (\ref{relesu}) proves the equality
\bal
(B^\pm)^{-1}_{gf}=\prod_{h=1}^{N_{\rm F}}\f{\Gamma(1+\tilde{m}_h-\tilde{m}_g)}{\Gamma(m_h-\tilde{m}_g)}
\f{\prod_{h\neq s}^{N_{\rm F}}\Gamma(-m_f+m_h)}{\prod_{h=1}^{N_{\rm F}}\Gamma(1+\tilde{m}_h-m_f)}\f{\pi e^{\mp i\pi  (m_f-\tilde{m}_g)}}{\sin\pi (m_f-\tilde{m}_g)}\, ,
\eal
which we checked numerically.

For the monodromy around $x=1$,
the formula (\ref{Psi-monodromy-SQED}) for $\Psi^{({\rm s})}_\lambda$ reduces, upon integration, to the one for the vortex partition functions $f^{({\rm s})}_g(x)$
\ba
x^{m_g}f^{({\rm s})}_g(x)\r \sum_{f}
\mathcal{M}_{gf}
x^{m_f}f^{({\rm s})}_f(x)\, ,
\ea
where the monodromy matrix is given by
\begin{equation}\label{M-cal-expression}
\begin{aligned}
(\mathcal{M}-{\rm id})_{gf} &=
-2i
\f{\prod_{h=1}^{N_{\rm F}}\sin \pi(m_f-\tilde{m}_h)}{\prod_{h\neq s}^{N_{\rm F}}\sin\pi (m_f-m_h)}
\\
&\qquad\qquad
\times
\Bigg(
\prod_{h=1}^{N_{\rm F}}\f{\Gamma(1-m_h+m_g)}{\Gamma(-\tilde{m}_h+m_g)}
\f{\Gamma(-\tilde{m}_h+m_f)}{\Gamma(1-m_h+m_f)}e^{-i\pi(m_h-\tilde{m}_h)}
\Bigg)
\,.
\end{aligned}
\end{equation}
The reference \cite{Gomis:2014eya} considered the monodromy along the contour in the direction opposite to ours (given in Figure \ref{figure:loop-x=1}) and obtained the matrix $M_1=B^+(B^-)^{-1}$.
Then the monodromy matrix $\mathcal{M}$ for our contour should coincide with $M_1^{-1}=B^-(B^+)^{-1}$.
Indeed we checked analytically for $N_\text{F}=2$ and numerically for $N_\text{F}=3,4$ that the right-hand side of (\ref{M-cal-expression}) equals $(B^{-}(B^+)^{-1})_{gf}-\delta_{fg}$.
Our computation by analytic continuation amounts to a general analytic proof of the equality.

\bibliography{refs}

\providecommand{\href}[2]{#2}\begingroup\raggedright\begin{thebibliography}{10}

\bibitem{Bachas:2001vj}
C.~Bachas, J.~de~Boer, R.~Dijkgraaf, and H.~Ooguri, ``{Permeable conformal
  walls and holography},''
  \href{http://dx.doi.org/10.1088/1126-6708/2002/06/027}{{\em JHEP} {\bfseries
  06} (2002) 027},
\href{http://arxiv.org/abs/hep-th/0111210}{{\ttfamily arXiv:hep-th/0111210
  [hep-th]}}.

\bibitem{Bak:2003jk}
D.~Bak, M.~Gutperle, and S.~Hirano, ``{A Dilatonic deformation of AdS(5) and
  its field theory dual},''
  \href{http://dx.doi.org/10.1088/1126-6708/2003/05/072}{{\em JHEP} {\bfseries
  05} (2003) 072},
\href{http://arxiv.org/abs/hep-th/0304129}{{\ttfamily arXiv:hep-th/0304129
  [hep-th]}}.

\bibitem{Clark:2004sb}
A.~B. Clark, D.~Z. Freedman, A.~Karch, and M.~Schnabl, ``{Dual of the Janus
  solution: An interface conformal field theory},''
  \href{http://dx.doi.org/10.1103/PhysRevD.71.066003}{{\em Phys. Rev.}
  {\bfseries D71} (2005) 066003},
\href{http://arxiv.org/abs/hep-th/0407073}{{\ttfamily arXiv:hep-th/0407073
  [hep-th]}}.

\bibitem{Brunner:2007ur}
I.~Brunner and D.~Roggenkamp, ``{Defects and bulk perturbations of boundary
  Landau-Ginzburg orbifolds},''
  \href{http://dx.doi.org/10.1088/1126-6708/2008/04/001}{{\em JHEP} {\bfseries
  04} (2008) 001},
\href{http://arxiv.org/abs/0712.0188}{{\ttfamily arXiv:0712.0188 [hep-th]}}.

\bibitem{Brunner:2008fa}
I.~Brunner, H.~Jockers, and D.~Roggenkamp, ``{Defects and D-Brane
  Monodromies},'' \href{http://dx.doi.org/10.4310/ATMP.2009.v13.n4.a4}{{\em
  Adv. Theor. Math. Phys.} {\bfseries 13} no.~4, (2009) 1077--1135},
\href{http://arxiv.org/abs/0806.4734}{{\ttfamily arXiv:0806.4734 [hep-th]}}.

\bibitem{Gaiotto:2009fs}
D.~Gaiotto, ``{Surface Operators in N = 2 4d Gauge Theories},''
  \href{http://dx.doi.org/10.1007/JHEP11(2012)090}{{\em JHEP} {\bfseries 11}
  (2012) 090},
\href{http://arxiv.org/abs/0911.1316}{{\ttfamily arXiv:0911.1316 [hep-th]}}.

\bibitem{Jockers:2012dk}
H.~Jockers, V.~Kumar, J.~M. Lapan, D.~R. Morrison, and M.~Romo, ``{Two-Sphere
  Partition Functions and Gromov-Witten Invariants},''
  \href{http://dx.doi.org/10.1007/s00220-013-1874-z}{{\em Commun. Math. Phys.}
  {\bfseries 325} (2014) 1139--1170},
\href{http://arxiv.org/abs/1208.6244}{{\ttfamily arXiv:1208.6244 [hep-th]}}.

\bibitem{Gomis:2012wy}
J.~Gomis and S.~Lee, ``{Exact Kahler Potential from Gauge Theory and Mirror
  Symmetry},'' \href{http://dx.doi.org/10.1007/JHEP04(2013)019}{{\em JHEP}
  {\bfseries 04} (2013) 019},
\href{http://arxiv.org/abs/1210.6022}{{\ttfamily arXiv:1210.6022 [hep-th]}}.

\bibitem{Gerchkovitz:2014gta}
E.~Gerchkovitz, J.~Gomis, and Z.~Komargodski, ``{Sphere Partition Functions and
  the Zamolodchikov Metric},''
  \href{http://dx.doi.org/10.1007/JHEP11(2014)001}{{\em JHEP} {\bfseries 11}
  (2014) 001},
\href{http://arxiv.org/abs/1405.7271}{{\ttfamily arXiv:1405.7271 [hep-th]}}.

\bibitem{Benini:2012ui}
F.~Benini and S.~Cremonesi, ``{Partition Functions of ${\mathcal{N}=(2,2)}$
  Gauge Theories on S$^{2}$ and Vortices},''
  \href{http://dx.doi.org/10.1007/s00220-014-2112-z}{{\em Commun. Math. Phys.}
  {\bfseries 334} no.~3, (2015) 1483--1527},
\href{http://arxiv.org/abs/1206.2356}{{\ttfamily arXiv:1206.2356 [hep-th]}}.

\bibitem{Doroud:2012xw}
N.~Doroud, J.~Gomis, B.~Le~Floch, and S.~Lee, ``{Exact Results in D=2
  Supersymmetric Gauge Theories},''
  \href{http://dx.doi.org/10.1007/JHEP05(2013)093}{{\em JHEP} {\bfseries 05}
  (2013) 093},
\href{http://arxiv.org/abs/1206.2606}{{\ttfamily arXiv:1206.2606 [hep-th]}}.

\bibitem{Bachas:2016bzn}
C.~Bachas and D.~Plencner, ``{Boundary Weyl anomaly of $ \mathcal{N} $ = (2, 2)
  superconformal models},''
  \href{http://dx.doi.org/10.1007/JHEP03(2017)034}{{\em JHEP} {\bfseries 03}
  (2017) 034},
\href{http://arxiv.org/abs/1612.06386}{{\ttfamily arXiv:1612.06386 [hep-th]}}.

\bibitem{Sugishita:2013jca}
S.~Sugishita and S.~Terashima, ``{Exact Results in Supersymmetric Field
  Theories on Manifolds with Boundaries},''
  \href{http://dx.doi.org/10.1007/JHEP11(2013)021}{{\em JHEP} {\bfseries 11}
  (2013) 021},
\href{http://arxiv.org/abs/1308.1973}{{\ttfamily arXiv:1308.1973 [hep-th]}}.

\bibitem{Honda:2013uca}
D.~Honda and T.~Okuda, ``{Exact results for boundaries and domain walls in 2d
  supersymmetric theories},''
  \href{http://dx.doi.org/10.1007/JHEP09(2015)140}{{\em JHEP} {\bfseries 09}
  (2015) 140},
\href{http://arxiv.org/abs/1308.2217}{{\ttfamily arXiv:1308.2217 [hep-th]}}.

\bibitem{Hori:2013ika}
K.~Hori and M.~Romo, ``{Exact Results In Two-Dimensional (2,2) Supersymmetric
  Gauge Theories With Boundary},''
\href{http://arxiv.org/abs/1308.2438}{{\ttfamily arXiv:1308.2438 [hep-th]}}.

\bibitem{Bachas:2013nxa}
C.~P. Bachas, I.~Brunner, M.~R. Douglas, and L.~Rastelli, ``{Calabi's diastasis
  as interface entropy},''
  \href{http://dx.doi.org/10.1103/PhysRevD.90.045004}{{\em Phys. Rev.}
  {\bfseries D90} no.~4, (2014) 045004},
\href{http://arxiv.org/abs/1311.2202}{{\ttfamily arXiv:1311.2202 [hep-th]}}.

\bibitem{MR0057000}
E.~Calabi, ``Isometric imbedding of complex manifolds,''
  \href{http://dx.doi.org/10.2307/1969817}{{\em Ann. of Math. (2)} {\bfseries
  58} (1953) 1--23}. \url{https://doi.org/10.2307/1969817}.

\bibitem{Cecotti:2013mba}
S.~Cecotti, D.~Gaiotto, and C.~Vafa, ``{$tt^*$ geometry in 3 and 4
  dimensions},'' \href{http://dx.doi.org/10.1007/JHEP05(2014)055}{{\em JHEP}
  {\bfseries 05} (2014) 055},
\href{http://arxiv.org/abs/1312.1008}{{\ttfamily arXiv:1312.1008 [hep-th]}}.

\bibitem{MR1354600}
A.~B. Givental, ``Homological geometry. {I}. {P}rojective hypersurfaces,''
  \href{http://dx.doi.org/10.1007/BF01671568}{{\em Selecta Math. (N.S.)}
  {\bfseries 1} no.~2, (1995) 325--345}.
  \url{https://doi.org/10.1007/BF01671568}.

\bibitem{1996alg.geom..3021G}
A.~B. {Givental}, ``{Equivariant Gromov - Witten Invariants},'' in {\em eprint
  arXiv:alg-geom/9603021}.
\newblock Mar., 1996.

\bibitem{MR1653024}
A.~Givental, ``A mirror theorem for toric complete intersections,'' in {\em
  Topological field theory, primitive forms and related topics ({K}yoto,
  1996)}, vol.~160 of {\em Progr. Math.}, pp.~141--175.
\newblock Birkh\"auser Boston, Boston, MA, 1998.
\newblock \href{http://arxiv.org/abs/alg-geom/9701016}{{\ttfamily
  arXiv:alg-geom/9701016 [alg-geom]}}.

\bibitem{Closset:2015rna}
C.~Closset, S.~Cremonesi, and D.~S. Park, ``{The equivariant A-twist and gauged
  linear sigma models on the two-sphere},''
  \href{http://dx.doi.org/10.1007/JHEP06(2015)076}{{\em JHEP} {\bfseries 06}
  (2015) 076},
\href{http://arxiv.org/abs/1504.06308}{{\ttfamily arXiv:1504.06308 [hep-th]}}.

\bibitem{Lerche:1989uy}
W.~Lerche, C.~Vafa, and N.~P. Warner, ``{Chiral Rings in N=2 Superconformal
  Theories},''
\href{http://dx.doi.org/10.1016/0550-3213(89)90474-4}{{\em Nucl. Phys.}
  {\bfseries B324} (1989) 427--474}.

\bibitem{Recknagel:1997sb}
A.~Recknagel and V.~Schomerus, ``{D-branes in Gepner models},''
  \href{http://dx.doi.org/10.1016/S0550-3213(98)00468-4}{{\em Nucl. Phys.}
  {\bfseries B531} (1998) 185--225},
\href{http://arxiv.org/abs/hep-th/9712186}{{\ttfamily arXiv:hep-th/9712186
  [hep-th]}}.

\bibitem{Seiberg:1993vc}
N.~Seiberg, ``{Naturalness versus supersymmetric nonrenormalization
  theorems},'' \href{http://dx.doi.org/10.1016/0370-2693(93)91541-T}{{\em Phys.
  Lett.} {\bfseries B318} (1993) 469--475},
\href{http://arxiv.org/abs/hep-ph/9309335}{{\ttfamily arXiv:hep-ph/9309335
  [hep-ph]}}.

\bibitem{Gaiotto:2008sd}
D.~Gaiotto and E.~Witten, ``{Janus Configurations, Chern-Simons Couplings, And
  The theta-Angle in N=4 Super Yang-Mills Theory},''
  \href{http://dx.doi.org/10.1007/JHEP06(2010)097}{{\em JHEP} {\bfseries 06}
  (2010) 097},
\href{http://arxiv.org/abs/0804.2907}{{\ttfamily arXiv:0804.2907 [hep-th]}}.

\bibitem{Cecotti:1991me}
S.~Cecotti and C.~Vafa, ``{Topological antitopological fusion},''
\href{http://dx.doi.org/10.1016/0550-3213(91)90021-O}{{\em Nucl. Phys.}
  {\bfseries B367} (1991) 359--461}.

\bibitem{Closset:2014pda}
C.~Closset and S.~Cremonesi, ``{Comments on $ \mathcal{N} $ = (2, 2)
  supersymmetry on two-manifolds},''
  \href{http://dx.doi.org/10.1007/JHEP07(2014)075}{{\em JHEP} {\bfseries 07}
  (2014) 075},
\href{http://arxiv.org/abs/1404.2636}{{\ttfamily arXiv:1404.2636 [hep-th]}}.

\bibitem{Okuda:2017rwo}
T.~Okuda, ``{Comments on supersymmetric renormalization in two-dimensional
  curved spacetime},'' \href{http://dx.doi.org/10.1007/JHEP12(2017)081}{{\em
  JHEP} {\bfseries 12} (2017) 081},
\href{http://arxiv.org/abs/1705.06118}{{\ttfamily arXiv:1705.06118 [hep-th]}}.

\bibitem{Knapp:2016rec}
J.~Knapp, M.~Romo, and E.~Scheidegger, ``{Hemisphere Partition Function and
  Analytic Continuation to the Conifold Point},''
  \href{http://dx.doi.org/10.4310/CNTP.2017.v11.n1.a3}{{\em Commun. Num. Theor.
  Phys.} {\bfseries 11} (2017) 73--164},
\href{http://arxiv.org/abs/1602.01382}{{\ttfamily arXiv:1602.01382 [hep-th]}}.

\bibitem{Erkinger:2017aaa}
D.~Erkinger and J.~Knapp, ``{Hemisphere Partition Function and Monodromy},''
  \href{http://dx.doi.org/10.1007/JHEP05(2017)150}{{\em JHEP} {\bfseries 05}
  (2017) 150},
\href{http://arxiv.org/abs/1704.00901}{{\ttfamily arXiv:1704.00901 [hep-th]}}.

\bibitem{MR717607}
R.~L. Bryant and P.~A. Griffiths, ``Some observations on the infinitesimal
  period relations for regular threefolds with trivial canonical bundle,'' in
  {\em Arithmetic and geometry, {V}ol. {II}}, vol.~36 of {\em Progr. Math.},
  pp.~77--102.
\newblock Birkh\"auser Boston, Boston, MA, 1983.

\bibitem{Strominger:1990pd}
A.~Strominger, ``{SPECIAL GEOMETRY},''
\href{http://dx.doi.org/10.1007/BF02096559}{{\em Commun. Math. Phys.}
  {\bfseries 133} (1990) 163--180}.

\bibitem{Freed:1997dp}
D.~S. Freed, ``{Special Kahler manifolds},''
  \href{http://dx.doi.org/10.1007/s002200050604}{{\em Commun. Math. Phys.}
  {\bfseries 203} (1999) 31--52},
\href{http://arxiv.org/abs/hep-th/9712042}{{\ttfamily arXiv:hep-th/9712042
  [hep-th]}}.

\bibitem{Horja-hypergeometric}
R.~P. {Horja}, ``{Hypergeometric functions and mirror symmetry in toric
  varieties},'' \href{http://arxiv.org/abs/math/9912109}{{\ttfamily
  arXiv:math/9912109 [math.AG]}}.

\bibitem{Bonelli:2013mma}
G.~Bonelli, A.~Sciarappa, A.~Tanzini, and P.~Vasko, ``{Vortex partition
  functions, wall crossing and equivariant Gromov-Witten invariants},''
  \href{http://dx.doi.org/10.1007/s00220-014-2193-8}{{\em Commun. Math. Phys.}
  {\bfseries 333} no.~2, (2015) 717--760},
\href{http://arxiv.org/abs/1307.5997}{{\ttfamily arXiv:1307.5997 [hep-th]}}.

\bibitem{Hosono:2000eb}
S.~Hosono, ``{Local mirror symmetry and type IIA monodromy of Calabi-Yau
  manifolds},'' \href{http://dx.doi.org/10.4310/ATMP.2000.v4.n2.a5}{{\em Adv.
  Theor. Math. Phys.} {\bfseries 4} (2000) 335--376},
\href{http://arxiv.org/abs/hep-th/0007071}{{\ttfamily arXiv:hep-th/0007071
  [hep-th]}}.

\bibitem{Jockers:2006sm}
H.~Jockers, ``{D-brane monodromies from a matrix-factorization perspective},''
  \href{http://dx.doi.org/10.1088/1126-6708/2007/02/006}{{\em JHEP} {\bfseries
  02} (2007) 006},
\href{http://arxiv.org/abs/hep-th/0612095}{{\ttfamily arXiv:hep-th/0612095
  [hep-th]}}.

\bibitem{Hosono:2004jp}
S.~Hosono, ``{Central charges, symplectic forms, and hypergeometric series in
  local mirror symmetry},''
\href{http://arxiv.org/abs/hep-th/0404043}{{\ttfamily arXiv:hep-th/0404043
  [hep-th]}}.

\bibitem{Candelas:1990rm}
P.~Candelas, X.~C. De~La~Ossa, P.~S. Green, and L.~Parkes, ``{A Pair of
  Calabi-Yau manifolds as an exactly soluble superconformal theory},''
  \href{http://dx.doi.org/10.1016/0550-3213(91)90292-6}{{\em Nucl. Phys.}
  {\bfseries B359} (1991) 21--74}.
[AMS/IP Stud. Adv. Math.9,31(1998)].

\bibitem{Hosono:1994ax}
S.~Hosono, A.~Klemm, S.~Theisen, and S.-T. Yau, ``{Mirror symmetry, mirror map
  and applications to complete intersection Calabi-Yau spaces},''
  \href{http://dx.doi.org/10.1016/0550-3213(94)00440-P}{{\em Nucl. Phys.}
  {\bfseries B433} (1995) 501--554},
  \href{http://arxiv.org/abs/hep-th/9406055}{{\ttfamily arXiv:hep-th/9406055
  [hep-th]}}.
[AMS/IP Stud. Adv. Math.1,545(1996)].

\bibitem{Gomis:2014eya}
J.~Gomis and B.~Le~Floch, ``{M2-brane surface operators and gauge theory
  dualities in Toda},'' \href{http://dx.doi.org/10.1007/JHEP04(2016)183}{{\em
  JHEP} {\bfseries 04} (2016) 183},
\href{http://arxiv.org/abs/1407.1852}{{\ttfamily arXiv:1407.1852 [hep-th]}}.

\bibitem{Hama:2012bg}
N.~Hama and K.~Hosomichi, ``{Seiberg-Witten Theories on Ellipsoids},''
  \href{http://dx.doi.org/10.1007/JHEP09(2012)033,
  10.1007/JHEP10(2012)051}{{\em JHEP} {\bfseries 09} (2012) 033},
  \href{http://arxiv.org/abs/1206.6359}{{\ttfamily arXiv:1206.6359 [hep-th]}}.
[Addendum: JHEP10,051(2012)].

\bibitem{Benini:2015noa}
F.~Benini and A.~Zaffaroni, ``{A topologically twisted index for
  three-dimensional supersymmetric theories},''
  \href{http://dx.doi.org/10.1007/JHEP07(2015)127}{{\em JHEP} {\bfseries 07}
  (2015) 127},
\href{http://arxiv.org/abs/1504.03698}{{\ttfamily arXiv:1504.03698 [hep-th]}}.

\bibitem{Witten:1988xj}
E.~Witten, ``{Topological Sigma Models},''
\href{http://dx.doi.org/10.1007/BF01466725}{{\em Commun. Math. Phys.}
  {\bfseries 118} (1988) 411}.

\bibitem{Morrison:1994fr}
D.~R. Morrison and M.~R. Plesser, ``{Summing the instantons: Quantum cohomology
  and mirror symmetry in toric varieties},''
  \href{http://dx.doi.org/10.1016/0550-3213(95)00061-V}{{\em Nucl. Phys.}
  {\bfseries B440} (1995) 279--354},
\href{http://arxiv.org/abs/hep-th/9412236}{{\ttfamily arXiv:hep-th/9412236
  [hep-th]}}.

\bibitem{Ueda:2016wfa}
K.~Ueda and Y.~Yoshida, ``{Equivariant A-twisted GLSM and Gromov--Witten
  invariants of CY 3-folds in Grassmannians},''
  \href{http://dx.doi.org/10.1007/JHEP09(2017)128}{{\em JHEP} {\bfseries 09}
  (2017) 128},
\href{http://arxiv.org/abs/1602.02487}{{\ttfamily arXiv:1602.02487 [hep-th]}}.

\bibitem{2016arXiv160708317K}
B.~{Kim}, J.~{Oh}, K.~{Ueda}, and Y.~{Yoshida}, ``{Residue mirror symmetry for
  Grassmannians},'' \href{http://arxiv.org/abs/1607.08317}{{\ttfamily
  arXiv:1607.08317 [math.AG]}}.

\bibitem{Gerhardus:2018zwb}
A.~Gerhardus, H.~Jockers, and U.~Ninad, ``{The Geometry of Gauged Linear Sigma
  Model Correlation Functions},''
\href{http://arxiv.org/abs/1803.10253}{{\ttfamily arXiv:1803.10253 [hep-th]}}.

\bibitem{Fourier-Mukai}
T.~Okuda. In progress.

\bibitem{Herbst:2008jq}
M.~Herbst, K.~Hori, and D.~Page, ``{Phases Of N=2 Theories In 1+1 Dimensions
  With Boundary},''
\href{http://arxiv.org/abs/0803.2045}{{\ttfamily arXiv:0803.2045 [hep-th]}}.

\bibitem{Dedushenko:2018tgx}
M.~Dedushenko, ``{Gluing II: Boundary Localization and Gluing Formulas},''
\href{http://arxiv.org/abs/1807.04278}{{\ttfamily arXiv:1807.04278 [hep-th]}}.

\bibitem{Howe:1987ba}
P.~S. Howe and G.~Papadopoulos, ``{N=2, D = 2 SUPERGEOMETRY},''
\href{http://dx.doi.org/10.1088/0264-9381/4/1/005}{{\em Class. Quant. Grav.}
  {\bfseries 4} (1987) 11--21}.

\bibitem{Grisaru:1994dm}
M.~T. Grisaru and M.~E. Wehlau, ``{Prepotentials for (2,2) supergravity},''
  \href{http://dx.doi.org/10.1142/S0217751X95000358}{{\em Int. J. Mod. Phys.}
  {\bfseries A10} (1995) 753--766},
\href{http://arxiv.org/abs/hep-th/9409043}{{\ttfamily arXiv:hep-th/9409043
  [hep-th]}}.

\bibitem{Grisaru:1995dr}
M.~T. Grisaru and M.~E. Wehlau, ``{Superspace measures, invariant actions, and
  component projection formulae for (2,2) supergravity},''
  \href{http://dx.doi.org/10.1016/0550-3213(95)00529-3}{{\em Nucl. Phys.}
  {\bfseries B457} (1995) 219--239},
\href{http://arxiv.org/abs/hep-th/9508139}{{\ttfamily arXiv:hep-th/9508139
  [hep-th]}}.

\bibitem{Gates:1995du}
S.~J. Gates, Jr., M.~T. Grisaru, and M.~E. Wehlau, ``{A Study of general 2-D,
  N=2 matter coupled to supergravity in superspace},''
  \href{http://dx.doi.org/10.1016/0550-3213(95)00648-6}{{\em Nucl. Phys.}
  {\bfseries B460} (1996) 579--614},
\href{http://arxiv.org/abs/hep-th/9509021}{{\ttfamily arXiv:hep-th/9509021
  [hep-th]}}.

\bibitem{Ketov:1996es}
S.~V. Ketov, ``{2-d, N=2 and N=4 supergravity and the Liouville theory in
  superspace},'' \href{http://dx.doi.org/10.1016/0370-2693(96)00332-2}{{\em
  Phys. Lett.} {\bfseries B377} (1996) 48--54},
\href{http://arxiv.org/abs/hep-th/9602038}{{\ttfamily arXiv:hep-th/9602038
  [hep-th]}}.

\bibitem{Shadchin:2006yz}
S.~Shadchin, ``{On F-term contribution to effective action},''
  \href{http://dx.doi.org/10.1088/1126-6708/2007/08/052}{{\em JHEP} {\bfseries
  08} (2007) 052},
\href{http://arxiv.org/abs/hep-th/0611278}{{\ttfamily arXiv:hep-th/0611278
  [hep-th]}}.

\end{thebibliography}\endgroup
 \end{document}